\documentclass[journal]{IEEEtran}
\makeatletter

\usepackage{blindtext}
\usepackage[T1]{fontenc}
\usepackage{textcomp}
\usepackage{graphicx}
\usepackage{amsmath}
\usepackage{amssymb}
\usepackage{bm}
\usepackage{mdwmath}
\usepackage{cases}
\usepackage{graphicx}
\usepackage{multirow}
\graphicspath{{figure/}{Biography/PDF/}}
\usepackage[dvipsnames]{xcolor}
\definecolor{myblack}{rgb}{0,0.4980,1} 
\definecolor{myred}{rgb}{0.8706,0.1608,0.0627} 
\usepackage[nosort,noadjust]{cite}
\usepackage{url}
\usepackage{tabularray}
\usepackage{subfigure}
\usepackage{cite}
\usepackage{amsmath,amssymb,amsfonts}
\usepackage{stfloats}
\usepackage{graphicx}
\usepackage{textcomp}
\usepackage{xcolor}
\usepackage{flushend}
\usepackage{float}
\usepackage{balance}
\usepackage{multirow}
\usepackage{algorithmicx}
\UseTblrLibrary{counter}

\usepackage{makecell}
\usepackage{siunitx}

\usepackage{hyperref}
\newcommand{\colorhypersetup}{\@ifpackageloaded{hyperref}{\hypersetup{%
	bookmarksopen=true,%
	bookmarksnumbered=true,%
	pdfpagemode={UseOutlines},
	pdfstartview={FitH},%
	colorlinks=true,%
	linkcolor={myred},%
    citecolor={orange}
}}{\empty}}
\newcommand{\blackhypersetup}{\@ifpackageloaded{hyperref}{\hypersetup{%
	bookmarksopen=true,%
	bookmarksnumbered=true,%
	pdfpagemode={UseOutlines},
	pdfstartview={FitH},%
	colorlinks=true,%
	allcolors={black}
}}{\empty}}
\blackhypersetup

\usepackage{mfirstuc-english}
\MFUhyphentrue
\usepackage[version3,description,IEEEtran]{eacro}
\acsetup{%
    pages/display=all,%
    pages/seq/use=false,%
    pages/name=true,%
    pages/fill={\quad},%
    make-links=true%
}
\DeclareAcronym{ULA}{
	short = ULA,
	long = uniform linear arrays}
\DeclareAcronym{AoA}{
	short = AoA,
	long = angle of arrival}
\DeclareAcronym{AoD}{
	short = AoD,
	long = angle of departure}
 \DeclareAcronym{DFRC}{
	short = DFRC,
	long = Dual Function Radar and Communication}
\DeclareAcronym{WGE}{
	short = WGE,
	long = weighted geometric mean}
\DeclareAcronym{SINR}{
	short = SINR,
	long = signal-to-interference-plus-noise ratio}
\DeclareAcronym{ISAC}{
	short = ISAC,
	long = Integrated Sensing and Communication}
\DeclareAcronym{IoV}{
	short = IoV,
	long = Internet of Vehicles}
\DeclareAcronym{6G}{
	short = 6G,
	long = sixth genearation}
\DeclareAcronym{AWGN}{
	short = AWGN,
	long = Additive Gaussian White Noise}
\DeclareAcronym{SNR}{
	short = SNR,
	long = signal-to-noise ratio}
\DeclareAcronym{RCC}{
	short = RCC,
	long = Radar-Communication Coexistence}
\DeclareAcronym{DDI}{
	short = DDI,
	long = double descent iteration}
 \DeclareAcronym{gNB}{
	short = gNB,
	long = next generation node B}

\usepackage{algorithm}
\usepackage[ruled, algo2e]{algorithm2e}
\usepackage[compatible]{algpseudocode}
\usepackage{enumitem}
\newcommand{\upperroman}[1]{\uppercase\expandafter{\romannumeral#1}}

\newcommand{\Rmnum}[1]{\expandafter\@slowromancap\romannumeral #1@}

\newcommand{\myunit}[1]{%
	\ifmmode
		\mathrm{#1}
	\else
		$ \mathrm{#1} $
	\fi}
\newcommand{\murm}{%
	\ifmmode
		\text{\textmu}
	\else
		\textmu
	\fi}

\newlength{\mysinglefigwidth}
\newlength{\mymultifigwidth}
\ifCLASSOPTIONonecolumn
	\AtBeginDocument{\setlength{\mysinglefigwidth}{0.5\linewidth}}
\else
	\AtBeginDocument{\setlength{\mysinglefigwidth}{\linewidth}}
\fi
\ifCLASSOPTIONonecolumn
	\AtBeginDocument{\setlength{\mymultifigwidth}{0.5\linewidth}}
\else
	\AtBeginDocument{\setlength{\mymultifigwidth}{\linewidth}}
\fi

\makeatother
\begin{document}
\ifCLASSOPTIONonecolumn
	\typeout{The onecolumn mode.}
	\title{\LARGE Title}
	\author{Author~1,~\IEEEmembership{Member,~IEEE}, and~Author~2,~\IEEEmembership{Fellow,~IEEE}
		\thanks{Manuscript received XXX, 2019; revised XXX.}
		\thanks{Author~1 is with the Intelligent Computing and Communications ($ \text{IC}^\text{2} $) Lab, Wireless Signal Processing and Networks (WSPN) Lab, Key Laboratory of Universal Wireless Communications, Ministry of Education, Beijing University of Posts and Telecommunications (BUPT), Beijing, 100876, China (E-mail: \textsf{XXX@bupt.edu.cn}).}
		\thanks{Author~2 is with the Intelligent Computing and Communications ($ \text{IC}^\text{2} $) Lab, Wireless Signal Processing and Networks (WSPN) Lab, Key Laboratory of Universal Wireless Communications, Ministry of Education, Beijing University of Posts and Telecommunications (BUPT), Beijing, 100876, China (E-mail: \textsf{XXX@bupt.edu.cn}).}
		\thanks{This work was supported by the National Natural Science Foundation of China (NSFC) under Grant No. XXX.}
	}
\else
	\typeout{The twocolumn mode.}
	\title{An Adaptive CSI Feedback Model Based on BiLSTM for Massive MIMO-OFDM Systems}
	\author{Hongrui~Shen, Long~Zhao,~\IEEEmembership{Member, IEEE}, Kan~Zheng,~\IEEEmembership{Fellow,~IEEE}, Yuhua~Cao, and Pingzhi~Fan,~\IEEEmembership{Fellow, IEEE}
        \thanks{Corresponding author: \textit{Long Zhao}.}
        \thanks{Hongrui Shen, Long Zhao are with Intelligent Computing and Communications Lab, Beijing University of Posts and Telecommunications (BUPT), Beijing, 100876, China (E-mail: shr1216@bupt.edu.cn; z-long@bupt.edu.cn).

        Kan Zheng is with the College of Electrical Engineering and Computer Sciences, Ningbo University, Ningbo, 315211, China (E-mail: zhengkan@nbu.edu.cn).
        
        Yuhua Cao is with the China Mobile Communication Research Institute (CMRI), Beijing 100053, China (E-mail: caoyuhua@chinamobile.com).
    
        Pingzhi Fan is with the Institute of Information Coding and Transmission Key Laboratory of Sichuan Province, Southwest Jiaotong University, Chengdu 610031, China (E-mail: pzfan@swjtu.edu.cn).}

	}
        
\fi

\ifCLASSOPTIONonecolumn
	\typeout{The onecolumn mode.}
\else
	\typeout{The twocolumn mode.}
\fi

\maketitle

\begin{abstract}
Deep learning (DL)-based channel state information (CSI) feedback has the potential to improve the recovery accuracy and reduce the feedback overhead in massive multiple-input multiple-output orthogonal frequency division multiplexing (MIMO-OFDM) systems. However, the length of input CSI and the number of feedback bits should be adjustable in different scenarios, which can not be efficiently achieved by the existing CSI feedback models. Therefore, an adaptive bidirectional long short-term memory network (ABLNet) for CSI feedback is first designed to process various input CSI lengths, where the number of feedback bits is in proportion to the CSI length. Then, to realize a more flexible feedback bit number, a feedback bit control unit (FBCU) module is proposed to control the output length of feedback bits. {\color{black}Based on which, a target feedback performance can be adaptively achieved by a designed bit number adjusting (BNA) algorithm.} Furthermore, a novel separate training approach is devised to solve the model protection problem that the UE and gNB are from different manufacturers. Experiments demonstrate that the proposed ABLNet with FBCU can fit for different input CSI lengths and feedback bit numbers; {\color{black}the CSI feedback performance can be stabilized by the BNA algorithm}; and the proposed separate training approach can maintain the feedback performance and reduce the complexity of feedback model.
\end{abstract}



\section{Introduction}

\acresetall

Massive multiple-input multiple-output (MIMO) technology has become the key technology of 5G mobile communication due to its high spectrum efficiency \cite{refer1,refer2}. To generate beamforming for downlink signal transmission, it is necessary to obtain timely and accurate downlink channel state information (CSI) at the next generation nodeB (gNB). In time division duplex (TDD) mode, the downlink CSI can be directly estimated from uplink pilot by using channel reciprocity \cite{add1}. While in frequency division duplex (FDD) mode, different frequency bands are employed for uplink and downlink, making it challenging to estimate the downlink CSI from uplink pilot. Consequently, the CSI feedback scheme for massive MIMO in FDD systems has become one of hot topics in recent years.
\par


{\color{black}{The traditional CSI feedback methods, such as compressive sensing (CS) algorithms \cite{refer5, refer6},  utilize the channel sparsity and orthogonal basis matrix to compress CSI and then feedback the low-dimension CSI to the gNB for reconstruction. However, the actual channel is not completely sparse on any orthogonal basis. Moreover, both the number of algorithm iterations and the feedback overhead increase dramatically with the increasing number of antennas in massive MIMO systems. Therefore, feasible CSI feedback methods for massive MIMO systems need to be further studied. 

Recently, artificial intelligence (AI) has been developed rapidly, including deep learning (DL) \cite{refer7, refer8, refer9} and reinforcement learning (RL) technology \cite{refer11, refer11_2, refer12}. In particular, wireless communication systems based on DL frameworks have attracted a lot of attention, which can efficiently achieve channel estimation \cite{refer14, refer15}, feedback  \cite{refer17, refer19, refer20, refer21, refer22, r7_new, refer27, refer28, refer29}, channel encoding\cite{refer23}, and signal detection \cite{refer24}. {\color{black}Specifically, the DL-based CSI feedback, including full channel CSI feedback and eigenvector-based CSI feedback, has been widely studied since it can provide higher recovery accuracy and lower feedback overhead simultaneously.}

{\color{black}For the full channel CSI feedback, a kind of CSI feedback architecture based on the autoencoder structure \cite{refer26}, termed as CsiNet, was first introduced in \cite{refer17}. {\color{black}Based on which, a series of feedback models were proposed to improve the feedback accuracy by optimizing the model structure or quantizing method of the compressed vector with continuous values \cite{refer19, refer20, refer21, refer22, r7_new}.} On the other hand, for the eigenvector-based CSI feedback discussed by 3rd generation partnership project (3GPP) \cite{3GPP}, fully connected layers and convolutional neural network (CNN) are adopted to design EVCsiNet \cite{refer27}. Moreover, the encoder of Transformer was used for both the encoder and decoder of CSI feedback model to compress and recover the eigenvectors \cite{refer28}. And in \cite{refer29}, authors proposed a lighter model, named MixerNet, based on MLP-Mixer structure \cite{refer30}.} 


These studies in \cite{refer17, refer19, refer20, refer21, refer22, r7_new, refer27, refer28, refer29} primarily focus on the design of DL-based models or quantization methods to improve the CSI feedback performance, {\color{black}they needs to train and store a large number of feedback models to handle different input CSI lengths and feedback bit numbers.} {\color{black}Therefore, for the full channel CSI feedback, the model with adaptive inputs or compression ratios (CRs) are studied. On the one hand, a DL-based feedback schemes only relying on  fully convolutional are proposed for CSI with different numbers of subcarriers and antennas, however realizing multiple CRs needs a list of model parameters \cite{r1_new, r6_new, r2_new}. On the other hand, a framework consisting of an encoder with multiple fully connected layers and multiple decoders is proposed to realize various CRs by leveraging optional output layer and corresponding decoder in \cite{refer32}, furthermore its variant simplifies multiple decoders into a multi-branches decoder \cite{r3_new}. However, both of them need more fully connected layers of encoder and decoders (or branches) to accommodate various CRs; {\color{black}meanwhile, the adaptive input model in \cite{r1_new, r6_new, r2_new} could not compatible with the multi-CRs model in \cite{refer32, r3_new}}. Additionally, the padding operation is taken to change the lengths of compressed vectors \cite{r4_new}; a classification model is designed to select the suitable CR for the CSI data with same size in the training stage, but can not be flexibly adjusted in the inference stage \cite{r5_new}; a quantization method with adjusted quantization bit number of each neuron output value is proposed for variable CRs without considering the quantization loss in the test stage \cite{r8_new}. In conclusion, the proposed adaptive schemes may not be sufficient for alterable input CSI lengths and feedback bit numbers simultaneously with one set of model parameters.}

Besides, most researchers do not consider the model protection problem \cite{add2, add3}, where the user equipment (UE) and gNB can not share the trained encoder and decoder model, including the model structure and parameters, when they belong to different manufactures in actual communication. Consequently, they need to unify their interface and train a pair of matched encoder and decoder separately. As a result, we are inspired to further investigate the CSI feedback model based on DL for these practical factors.



}} 

Therefore, this paper proposes an adaptive eigenvector-base CSI feedback model and designs a plug-in control unit for adapting to various input CSI lengths and feedback bit numbers simultaneously. {\color{black}Then, an adaptive bit number adjusting (BNA) algorithm is developed for satisfying the target feedback performance.} Moreover, a novel separate training approach is designed for CSI feedback when the UE and gNB are from different manufacturers. In summary, the main contributions of this paper are given as follows.

\begin{itemize}
\item[$\bullet$] {\color{black}To deal with different lengths of input CSI, an input-adaptive CSI feedback model based on autoencoder structure is designed by leveraging bidirectional long short-term memory (BiLSTM), referred to as ABLNet. By fully exploiting the property of LSTM for processing different lengths of input sequences, the proposed ABLNet is capable of compressing and recovering input CSI with different lengths and the number of feedback bits is in proportion to the corresponding length of input CSI.} 
\end{itemize}
\begin{itemize}
\item[$\bullet$] {\color{black}Then, a feedback bit control unit (FBCU) is proposed to further realize the adjustable number of feedback bits through discarding the ending part of codeword output by the encoder. Based on the adjustable characteristic of the proposed ABLNet with FBCU, a bit number adjusting (BNA) algorithm is designed to achieve a unified, instead of an average, target feedback performance for every input CSI with lower feedback bit number on average than before the adjustment.}
\end{itemize}
\begin{itemize}
\item[$\bullet$] {\color{black}In order to achieve the model protection between different manufacturers of encoder and decoder, the UE-first separate training method is adopted to decouple the training of the encoder and decoder. Moreover, a general decoder is utilized at the gNB to reconstruct CSI from different UEs rather than storing multiple decoders corresponding to multiple UE encoders. In this way, the UEs can adapt to different input CSI lengths and feedback bit numbers by utilizing the ABLNet with FBCU; and the general decoder can reduce the feedback model complexity.}
\end{itemize}

The rest of this paper is organized as follows. The massive MIMO orthogonal frequency division multiplexing (MIMO-OFDM) system model and DL-based CSI feedback model are described in Section \ref{section2}. {\color{black}Section \ref{section3} specifically describes the proposed ABLNet and FBCU for different lengths of input CSI and numbers of feedback bits, and introduces the BNA algorithm as well as studying the separate training method. Then, experiments are implemented to verify the advantage of proposed ABLNet with FBCU, the designed BNA algorithm and the separate training method in Section \ref{section4}.} Finally, Section \ref{section5} concludes this paper.

\section{System Model}
\label{section2}

This section firstly introduces massive MIMO-OFDM system model and then discusses the DL-based CSI feedback model in detail.

\begin{figure*}[!t]
	\centering
	\includegraphics[scale=0.7]{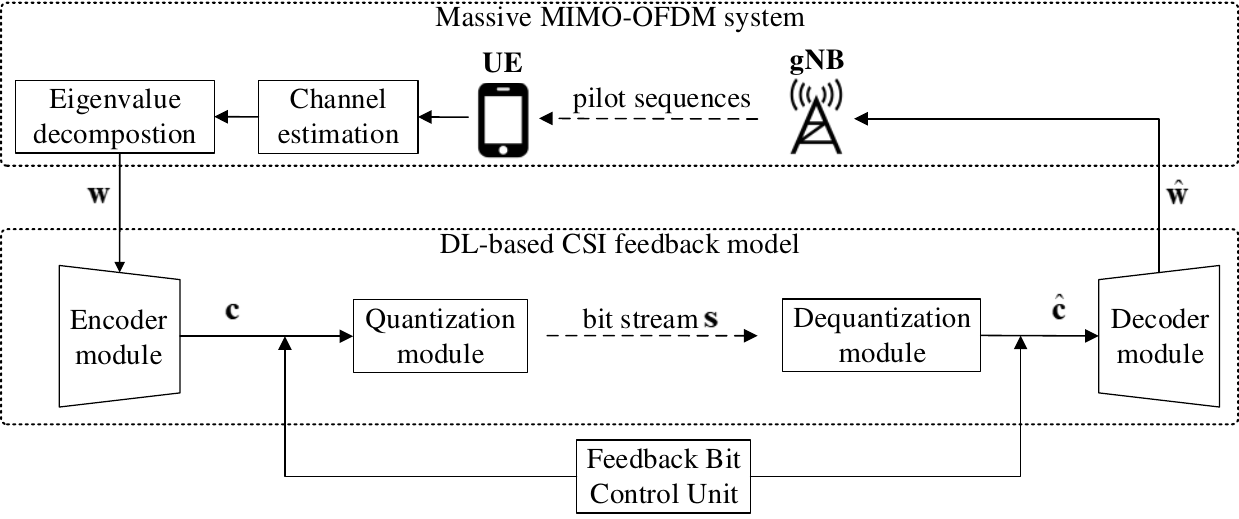}
	\caption{The massive MIMO-OFDM system model and adaptive CSI feedback framework.}
	\label{FIGURE1}
	\vspace{-0.1cm} 
\end{figure*}
\vspace{-0.1cm}
\subsection{Massive MIMO-OFDM System Model}
In a massive MIMO-OFDM system, $N_{\rm{T}}$ transmit antennas are deployed at the gNB and $N_{\rm{R}}$ receive antennas are deployed at each UE. As illustrated in Fig. \ref{FIGURE1}, by leveraging channel estimation at the UE \cite{refer33}, the downlink channels in frequency domain can be obtained based on the pilot sequences transmitted by the gNB. Then, the downlink channels are divided into $K$ subbands, where each subband consists of ${N_{{\rm{SC}}}}$ subcarriers and the subcarriers in the same subband employ the same beamformer at the gNB in order to reduce the system complexity. The common beamformer in each subband can be obtained by the following method in order to maintain the performance of each subcarrier.

Assume that the downlink channel of the \textit{n}th subcarrier in the \textit{k}th subband is denoted as ${{\bf{H}}_{kn}} \in {{\mathbb{C}}^{{N_{\rm{R}}} \times {N_{\rm{T}}}}}$ $\left( {1 \le k \le K, 1 \le n \le {N_{{\rm{SC}}}}} \right)$. Then, the correlation matrix of the channel of each subcarrier in the \textit{k}th subband can be calculated as ${{\bf{H}}_{kn}^{\rm{H}}{{\bf{H}}_{kn}}}$, and the average correlation matrix of the \textit{k}th subband can be written as
\begin{align}
	{{\bf{R}}_k} = \frac{1}{{{N_{{\rm{sc}}}}}}\sum\limits_{n = 1}^{{N_{{\rm{sc}}}}} {{\bf{H}}_{kn}^{\rm{H}}{{\bf{H}}_{kn}}}.
\end{align}
\par

Based on eigenvalue decomposition (EVD), the feedback eigenvector of the \textit{k}th subband can be calculated by
\begin{flalign} 
	{{\bf{R}}_k}{{\bf{w}}_k} = {\lambda _k}{{\bf{w}}_k}\label{eq2},
\end{flalign}
where ${\lambda _k}$ and ${{\bf{w}}_k} = {\left[ {{w_{k1}},{w_{k2}}, \cdots ,{w_{k{N_{\rm{T}}}}}} \right]^{\rm{T}}} \in {{\mathbb{C}}^{{N_{\rm{T}}} \times 1}}$ represent the maximum eigenvalue and corresponding eigenvector of matrix ${{\bf{R}}_k}$ in the \textit{k}th subband. From (\ref{eq2}), ${{\bf{w}}_k}$ is a complex vector; to fit for the processing of universal neural network, the corresponding real-valued eigenvector is given by
\begin{flalign*} 
	&{{\bf{\tilde w}}_k} =& 
\end{flalign*}
\vspace{-0.8cm}
\begin{flalign}
	{\left[ {{\mathop{\rm Re}\nolimits} \left\{ {{w_{k1}}} \right\},{\mathop{\rm Im}\nolimits} \left\{ {{w_{k1}}} \right\}, \cdots ,{\mathop{\rm Re}\nolimits} \left\{ {{w_{k{N_{\rm{T}}}}}} \right\},{\mathop{\rm Im}\nolimits} \left\{ {{w_{k{N_{\rm{T}}}}}} \right\}} \right]^{\rm{T}}},
\end{flalign}
where ${{\bf{\tilde w}}_k} \in {{\mathbb{R}}^{2{N_{\rm{T}}} \times 1}}$ and ${\mathop{\rm Re}\nolimits} \left\{  \cdot  \right\}$ and ${\mathop{\rm Im}\nolimits} \left\{  \cdot  \right\}$ represent the real value and the imaginary value of ${w_{kn}}\left( {1 \le n \le {N_{\rm{T}}}} \right)$, respectively. Therefore, the joint real eigenvector ${\bf{w}}$ of the $K$ subbands can be written as
\begin{flalign} 
	{\bf{w}} = {\left[ {{{{\bf{\tilde w}}}_1},{{{\bf{\tilde w}}}_2}, \cdot  \cdot  \cdot ,{{{\bf{\tilde w}}}_K}} \right]^{\rm{T}}} \in {{\mathbb{R}}^{K \times 2{N_{\rm{T}}}}}.
\end{flalign}

Moreover, CSI eigenvector with different lengths means that different CSI joint eigenvectors have different number of subbands, i.e., $K$.

\subsection{DL-Based CSI Feedback Model}

Since the CSI feedback scheme is comparable to the overall workflow of the autoencoder, the autoencoder architecture can be employed to design the DL-based CSI feedback framework. As shown in Fig. \ref{FIGURE1}, an adaptive CSI feedback model is proposed to adapt to the input CSI eigenvector with different lengths and various numbers of feedback bits. 
		
Among the feedback model, the UE first obtains the CSI eigenvector ${\bf{w}}$ and compresses it to codeword vector ${\bf{c}}$ by the encoder, which is denoted as
\begin{flalign} 
	{\bf{c}} = {f_{{{\bf{\theta }}_{\rm{E}}}}}\left( {{\bf{w}},K} \right)\label{eq5},
\end{flalign}
where ${f_{{{\bf{\theta }}_{\rm{E}}}}}\left(  \cdot  \right)$ is the encoder function with parameter set ${{\bf{\theta }}_{\rm{E}}}$. And the length $m$ of codeword vector ${\bf{c}}$ changes proportionally with $K$.

Then, a FBCU module is applied to make the codeword vector ${\bf{c}}$ changeable based on the capacity requirements of feedback channel. If the number $q$ of quantization bit is fixed, the length $Q$ of feedback bits obtained after quantization is also changeable according to the length $n\left( { \le m} \right)$ of ${\bf{c}}$. Therefore, if the FBCU module is applied, (\ref{eq5}) becomes
\begin{flalign} 
	{\bf{c}} = {f_{{{\bf{\theta }}_{\rm{E}}}}}\left( {{\bf{w}},K,n} \right).
\end{flalign}

{\color{black}After passing through the quantization and dequantization module, the obtained feedback bit stream ${\bf{s}}$ and recovered codeword vector ${\bf{\hat c}}$ can be respectively expressed as
\begin{flalign} 
	{\bf{s}} = {f_{{\rm{quan}}}}\left( {\bf{c}} \right),  {\bf{\hat c}} = {f_{{\rm{dequan}}}}\left( {\bf{s}} \right), 
\end{flalign}
where ${f_{{\rm{quan}}}}\left(  \cdot  \right)$ and ${f_{{\rm{dequan}}}}\left(  \cdot  \right)$ are the quantization and dequantization functions, respectively.} 

Finally, the gNB reconstructs the CSI eigenvector ${\bf{\hat w}}$ with the decoder module, which is denoted as
\begin{flalign} 
	{\bf{\hat w}} = {f_{{{\bf{\theta }}_{\rm{D}}}}}\left( {{\bf{\hat c}},K} \right),
\end{flalign}
where ${f_{{{\bf{\theta }}_{\rm{D}}}}}\left(  \cdot  \right)$ is the decoder function with parameter set ${{\bf{\theta }}_{\rm{D}}}$.

Usually, the square of generalized cosine similarity (SGCS) is taken to evaluate the recovery performance of eigenvector-based CSI feedback. 
{\color{black}The average SGCS $\rho$ between the original joint eigenvector ${\bf{w}}$ and the recovered joint eigenvector ${\bf{\hat w}}$ with \textit{K} subbands can be written as
\begin{flalign} 
	{\rho }\left( {{{\bf{\theta }}_{\rm{E}}},{{\bf{\theta }}_{\rm{D}}}} \right) = \frac{1}{K}\sum\limits_{k = 1}^K {{\left( {\frac{{\left\| {{\bf{w}}_k^{\rm{H}}{{{\bf{\hat w}}}_k}} \right\|}}{{\left\| {{{\bf{w}}_k}} \right\|\left\| {{{{\bf{\hat w}}}_k}} \right\|}}} \right)^2}}. \label{eq9}
\end{flalign}}
\par
{\color{black}Then, one objective of the paper is to optimize the model weight parameter set ${{\bf{\Theta }}_1} = \left\{ {{{\bf{\theta }}_{\rm{E}}},{{\bf{\theta }}_{\rm{D}}}} \right\}$ to make the feedback performance ${\rho}$ as high as possible, as most existing studies \cite{refer17, refer19, refer20, refer21, refer22, r7_new, refer27, refer28, refer29} have done.} On the other hand, for the proposed separate training with multiple UEs and a single gNB, the objective is to optimize the weight parameter set ${{\bf{\Theta }}_2} = \left\{ {{{\bf{\theta }}_{{\rm{E1}}}},{{\bf{\theta }}_{{\rm{E2}}}}, \cdots ,{{\bf{\theta }}_{{\rm{E}}N}},{{\bf{\theta }}_{\rm{D}}}} \right\}$ to make the feedback performance ${\rho}$ of all $N$ UEs as high as possible, i.e.,
\begin{flalign} 
	{{\bf{\Theta }}_2} = \mathop {\arg \max }\limits_{{{\bf{\Theta }}_2}} \frac{1}{N}\sum\limits_{n = 1}^N {\rho _n} \left( {{{\bf{\theta }}_{{\rm{E}}n}},{{\bf{\theta }}_{\rm{D}}}} \right),
\end{flalign}
where ${{\bf{\theta }}_{{\rm{E}}n}}$ and $\rho _n\left( {1 \le n \le N} \right)$ represent the encoder parameters set and the SGCS of the \textit{n}th UE, respectively.
\par

\section{Adaptive CSI Feedback Model}
\label{section3}

The adaptive CSI feedback model based on BiLSTM structure, named ABLNet, is first introduced in this section. The role of ABLNet includes adapting to different input lengths of  CSI eigenvectors and ensuring the recovery accuracy. Then, the FBCU module is designed to make the feedback bit number of ABLNet model more adjustable. {\color{black}Based on which, the BNA algorithm is introduced to achieve the target SGCS for every input CSI by adjusting the number of feedback bits.} In addition, a novel separate training approach is proposed to enable the CSI feedback between multiple UEs and a single gNB that come from different manufacturers.
\vspace{-0.1cm}

\begin{figure*}[!t]
	\centering
	\includegraphics[scale=0.56]{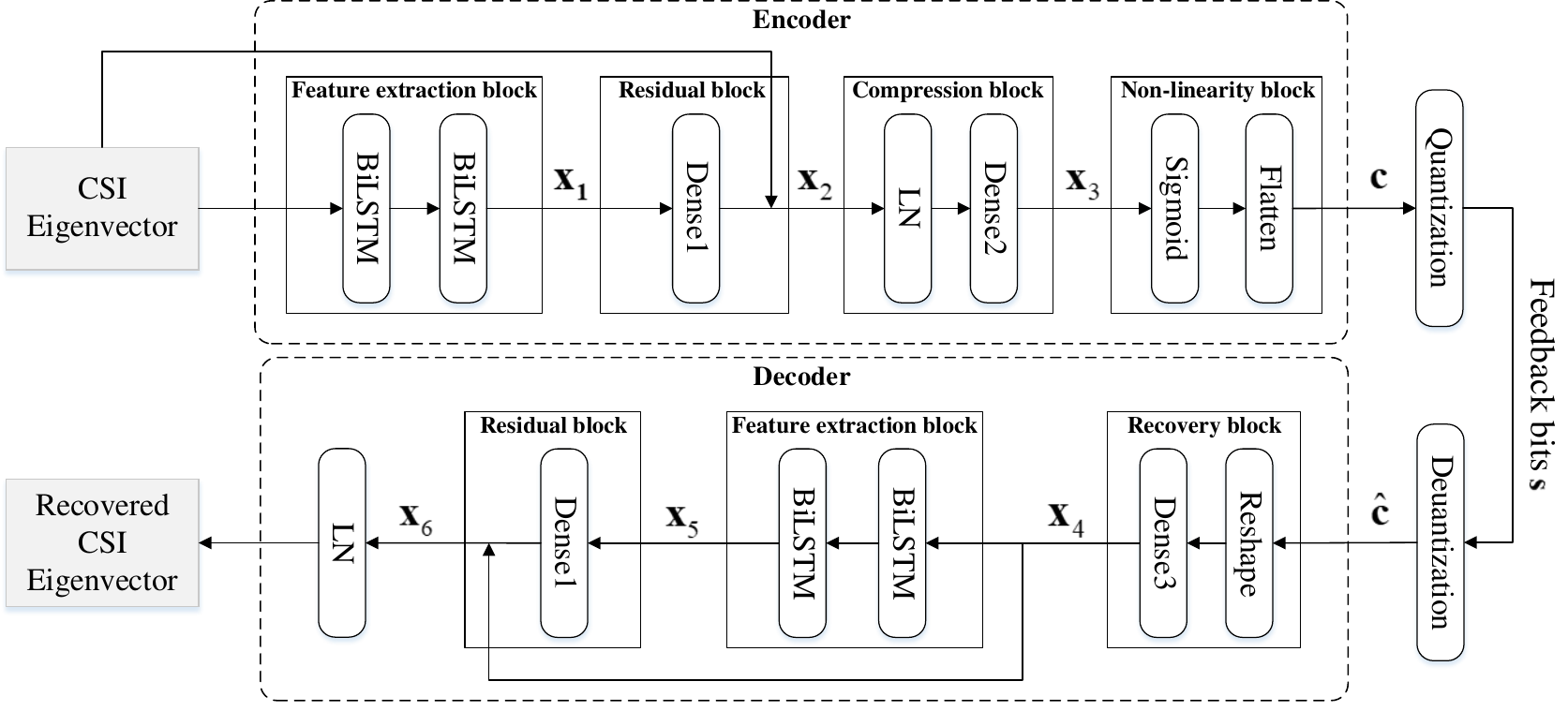}
	\caption{Proposed ABLNet architecture.}
	\label{FIGURE2}
\end{figure*}

\subsection{Architecture of ABLNet}

{\color{black}{By fully considering the characteristics of input CSI eigenvector and the aim of dealing with different subband numbers, the designed framework of ABLNet is illustrated in Fig. \ref{FIGURE2}, which contains encoder and decoder and both are described as follows.}}

\textbf{Encoder}. The input CSI eigenvector  sequentially passes through four types of blocks and the specific design of each block is described as follows.

\vspace{0.1cm}
\subsubsection{Feature extraction block} 
\begin{figure}[!t]
	\centering
	\includegraphics[width=\mysinglefigwidth]{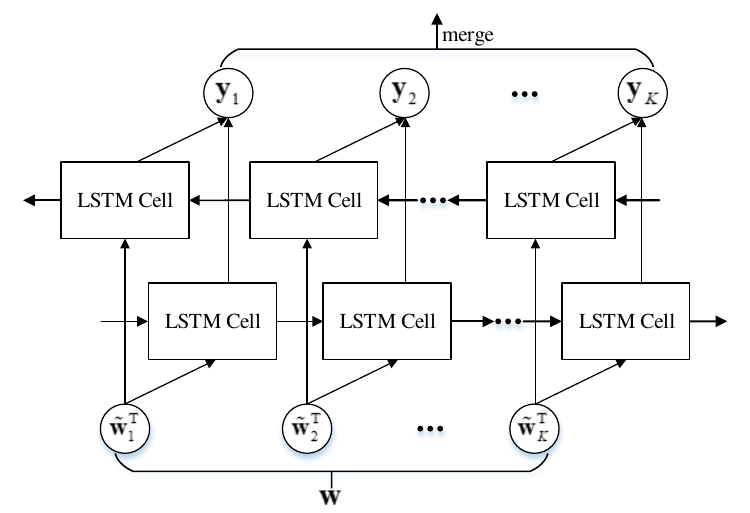}
	\vspace{-0.5cm}
	\caption{BiLSTM structure.}
	\label{FIGURE12}
	\vspace{-0.4cm} 
\end{figure}

The block consists of two BiLSTM layers, which is responsible for extracting the feature of input CSI eigenvector. The detailed BiLSTM structure is shown in Fig. \ref{FIGURE12}, where all subband eigenvectors ${\bf{\tilde w}}_1^{\rm{T}},{\bf{\tilde w}}_2^{\rm{T}}, \cdots ,{\bf{\tilde w}}_K^{\rm{T}}$ are regarded as a sequence. Each subband eigenvector is input to a LSTM cell and is extracted feature vector by using features of preceding and following subband eigenvectors. Then, the $K$ output feature vectors ${{\bf{y }}_1},{{\bf{y }}_2}, \cdots ,{{\bf{y }}_K}$ of ${\bf{\tilde w}}_1^{\rm{T}},{\bf{\tilde w}}_2^{\rm{T}}, \cdots ,{\bf{\tilde w}}_K^{\rm{T}}$ are merged to be transmitted to the next BiLSTM layer. After passing through this block, {\color{black}the output feature vector can be expressed as ${{\bf{x}}_1} = {\rm{BiLSTM}}\left( {{\rm{BiLSTM}}\left( {\bf{w}} \right)} \right)$.}
\vspace{0.1cm}
\subsubsection{Residual block} The residual block has a fully connected layer, i.e., Dense1, and transforms the output of feature extraction block into ${\rm{Dense1}}\left( {{{\bf{x}}_1}} \right)$ for further processing. Moreover, the residual structure \cite{refer34} is introduced to accelerate model convergence and improve the model performance. The original input eigenvector ${\bf{w}}$ is maintained via the shortcut branch and {\color{black}so the final output feature vector of the residual block can be written as ${{\bf{x}}_2} = {\rm{Dense1}}\left( {{{\bf{x}}_1}} \right) + {\bf{w}}$.}
\vspace{0.1cm}

\subsubsection{Compression block} Data compression is achieved in this block. Firstly, layer normalization (LN) operation is carried out to unify the variation range of the extracted feature vector ${{\bf{x}}_2}$ and help network training. Then, a fully connected layer, Dense2, is employed to compress the output ${\rm{LN}}\left( {{{\bf{x}}_2}} \right)$ and {\color{black}therefore the output of whole compression block can be expressed as ${{\bf{x}}_3} = {\rm{Dense2}}\left( {{\rm{LN}}\left( {{{\bf{x}}_2}} \right)} \right)$.}
\vspace{0.1cm}
\subsubsection{Non-linearity block} In order to utilize the non-linearity to improve the expressiveness of the model, the Sigmoid function is first adopted in the block. And then, the flatten operation is employed to transform its output ${{\rm{Sigmoid}}\left( {{{\bf{x}}_3}} \right)}$ {\color{black}into the codeword vector ${\bf{c}}$, which can be expressed as ${\rm{Flatten}}\left( {{\rm{Sigmoid}}\left( {{{\bf{x}}_3}} \right)} \right)$.}

Finally, the obtained codeword vector ${\bf{c}}$ is sent to the quantization layer and is quantized into the feedback bit stream ${\bf{s}}$. In this paper, the uniform quantization method is adopted to quantize the compressed codeword. Moreover, some other quantization methods introduced in \cite{refer29} and \cite{refer35}, such as non-uniform quantization and vector quantization, can also be used to improve the CSI feedback performance.



\textbf{Decoder}. At the gNB, the dequantization layer is firstly used to transform the received bit stream ${\bf{s}}$ into the float codeword vector ${\bf{\hat c}}$, which is the inverse operation of the quantization layer at the UE. Then, the codeword vector ${\bf{\hat c}}$ passes through the recovery block, feature extraction block and residual block in sequence.
\vspace{0.1cm}
\setcounter{subsubsection}{0}
\subsubsection{Recovery block} The input codeword vector ${\bf{\hat c}}$ is reshaped and then passes through the fully connected layer, i.e., Dense3, and {\color{black}the output vector can be written as ${{\bf{x}}_4} = {\rm{Dense3}}\left( {{\rm{Reshape}}\left( {{\bf{\hat c}}} \right)} \right)$.}

The output vector ${{\bf{x}}_4}$ has the same size as the original input eigenvector ${\bf{w}}$, while it does not represent the final recovered CSI eigenvector and further processing is needed.
\vspace{0.1cm}
\subsubsection{Feature extraction block \& Residual block} Similar to the encoder, the feature extraction block and residual block are used to extract the data feature. The input vector ${{\bf{x}}_4}$ passes through the feature extraction block and {\color{black}the obtained output feature can be expressed as ${{\bf{x}}_5} = {\rm{BiLSTM}}\left( {{\rm{BiLSTM}}\left( {{{\bf{x}}_4}} \right)} \right)$.}

Then, the feature ${{\bf{x}}_5}$ is fed into the residual block and also the residual structure is adopted to {\color{black}obtain the output as ${{\bf{x}}_6} = {\rm{Dense1}}\left( {{{\bf{x}}_5}} \right) + {{\bf{x}}_4}$.}

After that, the LN operation is adopted to get the final recovered CSI eigenvector ${\bf{\hat w}}$, {\color{black}which can be written as ${\bf{\hat w}} = {\rm{LN}}\left( {{{\bf{x}}_6}} \right)$.}

\begin{figure*}[t]
	\centering
	\includegraphics[scale=.9]{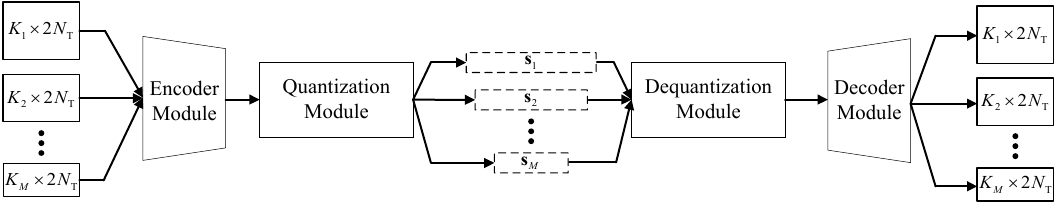}
	\caption{CSI feedback with different lengths of subband eigenvectors.}
	\label{FIGURE3}
\end{figure*}

\subsection{ABLNet for Adjustable Subband Number} 

\subsubsection{Deal with different subband numbers} {\color{black}Generally, DL-based models have the fixed-length input interface. As a result, in order to satisfy the fixed input length of the proposed ABLNet, the input CSI eigenvectors with different number of subbands need to be unified to the maximum number of subbands through the approach of padding 0. Meanwhile, the BiLSTM layer only processes the non-zero input part during the feature extraction and the zero-padding part is automatically ignored. Therefore, on the one hand, every subband eigenvector of each kind of subband number can be further compressed and quantified through the ABLNet after padding operation; on the other hand, the length $Q$ of feedback bit streams is in proportion to the number $K$ of CSI eigenvector subbands.}



{\color{black}In this way, when the ABLNet model receives CSI eigenvectors with different subband numbers in different scenarios, the feedback overhead can also change with the subband number.} As shown in Fig. \ref{FIGURE3}, CSI eigenvectors with subband number ${K_1},{K_2}, \cdots ,{K_M}$ are input to the model, and after compression and quantization, the lengths ${Q_1},{Q_2}, \cdots ,{Q_M}$ of feedback bits ${{\bf{s}}_1},{{\bf{s}}_2}, \cdots ,{{\bf{s}}_M}$ are proportional with the corresponding subband number of CSI eigenvectors, i.e.,
\begin{flalign} 
	\frac{{{Q_1}}}{{{K_1}}} = \frac{{{Q_2}}}{{{K_2}}} =  \cdots  = \frac{{{Q_M}}}{{{K_M}}}.
\end{flalign}



\subsubsection{Loss function} As mentioned in Section \ref{section2}, SGCS is used as the loss function of the proposed ABLNet model, which can be formulated as
\begin{flalign}
	{L_1} = {\sum\limits_{i = 1}^M {\sum\limits_{k = 1}^{{K_i}} {\frac{{{\mu _i}}}{{{K_i}}}\left( {\frac{{\left\| {{\bf{w}}_k^{\rm{H}}{{{\bf{\hat w}}}_k}} \right\|}}{{\left\| {{{\bf{w}}_k}} \right\|\left\| {{{{\bf{\hat w}}}_k}} \right\|}}} \right)} } ^2}\label{eq16},
\end{flalign}
where ${K_i} \in \left\{ {{K_1},{K_2}, \cdots ,{K_M}} \right\}$ is the number of CSI eigenvector subbands and ${{\mu _i}}$ is the weight coefficient of SGCS of CSI eigenvector with subband number ${K_i}$. In this paper, when CSI eigenvectors with multiple subband numbers are trained at the same time, equal weight coefficient ${\mu _i} = {1 \mathord{\left/
		{\vphantom {1 M}} \right.
		\kern-\nulldelimiterspace} M}$ is assumed for simplification. 
However, when the ABLNet is trained for a fixed subband number ${K_j}$, the loss function can be set as ${\mu _i} = 1\left( {i = j} \right)$ and ${\mu _i} = 0\left( {i \ne j} \right)$.

\subsection{FBCU for Adjustable Feedback Bit Number}
Since the encoder output length $m$ of the codeword vector ${\bf{c}}$ and the number $q$ of quantization bit for each float number in the output codeword vector ${\bf{c}}$ are predetermined, the CSI feedback bit stream ${\bf{s}}$ obtained by the UE has a constant length $Q$, which can be calculated as
\begin{flalign} 
	Q\left( m \right) = m \times q.
\end{flalign}

While in the actual communication systems, different scenarios or applications may require changeable feedback overhead, even for the same length of input eigenvector. {\color{black}And every subband CSI eigenvector can report different length of feedback information.} Therefore, as shown in Fig. \ref{FIGURE4}, a CSI feedback scheme with FBCU that adapts to different numbers of feedback bits is proposed to improve the generalization and practicability of the CSI feedback model.

\subsubsection{FBCU principle} 
\begin{figure*}[t]
	\centering
	\includegraphics[scale=0.9]{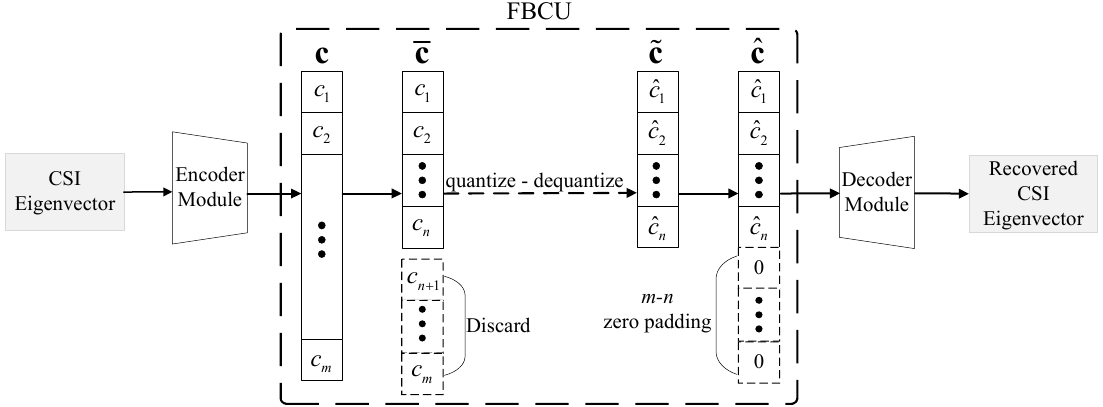}
	\caption{\rmfamily{CSI feedback scheme with FBCU.}}
	\label{FIGURE4}
	\vspace{0.3cm}
\end{figure*}

The FBCU is applied for CSI feedback model to change the length of the compressed vector output by the encoder and further change the number of feedback bits. Assuming the output codeword vector of the encoder is ${\bf{c}} = \left[ {{c_1},{c_2}, \cdots ,{c_m}} \right]$ in Fig. \ref{FIGURE4}, the FBCU can maintain the codeword vector with length $n$, i.e., ${\bf{\bar c}} = \left[ {{c_1},{c_2}, \cdots ,{c_n}} \right]$, according to the feedback overhead requirements and directly discard the rest part of the codeword vector, i.e., $\left[ {{c_{n + 1}},{c_{n + 2}}, \cdots ,{c_m}} \right]$. Then, the remaining codeword vector ${\bf{\bar c}}$ is quantized and fed back to the gNB through the feedback link. Due to the fact that the length $n$ is variable, the length $Q$ of feedback bit stream ${\bf{s}}$ obtained by quantization also changes with $n$, which can be written as
\begin{flalign} 
	Q\left( n \right) = n \times q.
\end{flalign}

At the gNB, after passing the dequantization layer, the dequantized codeword vector ${\bf{\tilde c}} = \left[ {{{\hat c}_1},{{\hat c}_2}, \cdots ,{{\hat c}_n}} \right]$ is obtained. Similar to the input of the encoder, the input dimension of the dequantized codeword vector to the decoder should also be fixed and the decoder can not directly deal with length-variable codeword vector. Therefore, the FBCU should unify the dequantized codeword vector ${\bf{\tilde c}}$ with different lengths $n$ into the codeword vector with the maximum length $m$ by the approach of padding zero, i.e., ${\bf{\hat c}} = \left[ {{{\hat c}_1},{{\hat c}_2}, \cdots ,{{\hat c}_n},   0,0 \cdots ,0} \right]$. Finally, the decoder reconstructs the CSI eigenvector ${\bf{\hat w}}$ from the zero-padded codeword vector ${\bf{\hat c}}$.

\subsubsection{ABLNet with FBCU} Combining FBCU and ABLNet can realize changeable input length of CSI eigenvector and feedback bit number by only one pair of encoder and decoder. Therefore, the ABLNet with FBCU can improve the generalization and practicability of CSI feedback model.  

To train the proposed ABLNet with FBCU, the length set ${{\bf{N}}_i}\left( {1 \le i \le M} \right)$ of codeword vector ${{\bf{\bar c}}}$ for the CSI eigenvectors with subband number ${K_i}$ should be first determined. Then, during the training process, a specific length ${n_i} \in {{\bf{N}}_i}$ is randomly selected in each training epoch. Once the length $n_i$ is fixed, the FBCU discards the last $m-n_i$ elements of the codeword vector ${\bf{c}}$ and pad zeros to the dequantized and truncated codeword vector ${\bf{\tilde c}}$ to guarantee the same input length $m$ of the codeword vector ${\bf{\hat c}}$ for the decoder. In this way, the SGCS can still be used as the loss function, which is the same as (\ref{eq16}). 

Moreover, since the FBCU only discards part of the compressed codeword vector ${\bf{c}}$ to change the length of feedback bit stream, it can also be applied to any DL-based CSI feedback model mentioned above \cite{refer17, refer19, refer20, refer21, refer22, r7_new, refer27, refer28, refer29} to realize the adjustable number of feedback bits.

{\color{black}\subsection{BNA Algorithm for Target SGCS}

\subsubsection{Purpose of BNA algorithm} Generally, when the CSI feedback performance approaches one target SGCS ${\rho _{\rm{t}}}$, further improving the SGCS will increase little communication performance and meanwhile consume a lot of feedback overhead. On the other hand, for the varying input CSI eigenvectors, the SGCS performance is different with a fixed number of feedback bits. Therefore, by utilizing the adjustable characteristic of feedback bit numbers of ABLNet with FBCU, a BNA algorithm is developed to increase or decrease the number $Q$ of feedback bits to make the SGCS performance $\rho$ stabilize at a reasonable target SGCS ${\rho _{\rm{t}}}$ for every feedback eigenvector. 
}


{\color{black}\subsubsection{Description of BNA algorithm} Assume that the minimum and maximum number of feedback bits supported by ABLNet with FBCU are ${{Q_{\rm{min}}}}$ and ${{Q_{\rm{max}}}}$, respectively; given the initial feedback bit number ${{Q_{\rm{i}}}}$, the corresponding SGCS performance ${{\rho_{\rm{i}}}}$, the target SGCS ${\rho _{\rm{t}}}$ and the tolerance error $\varepsilon=0.01$, the BNA algorithm is shown in Algorithm \ref{algorithm1}.


In Algorithm \ref{algorithm1}, according to the relation of $\left| {{\rho _{\rm{i}}} - {\rho _{\rm{t}}}} \right|$ and $\varepsilon$, the \textit{Binary\_Search} method is taken to find the appropriate feedback bit number ${{Q_{\rm{a}}}}$, making the corresponding feedback performance ${\rho _{\rm{a}}}$ within the tolerance error $\varepsilon$ of ${\rho _{\rm{t}}}$. Because the SGCS performance $\rho$ increases with the feedback bit number $Q$, taking the binary search strategy can improve the adjustment efficiency. Finally, both ${{Q_{\rm{a}}}}$ and ${\rho _{\rm{a}}}$ are the output of the BNA algorithm as the final adjusted feedback bit number and SGCS performance, respectively.

Moreover, if the ${\rho _{\rm{a}}}$ has reached the error $\varepsilon$ range of ${\rho _{\rm{t}}}$, but the adjusted feedback bit number ${Q _{\rm{a}}}$ is lower than ${Q _{\rm{min}}}$ or higher than ${Q _{\rm{max}}}$, the BNA algorithm takes the ${Q _{\rm{min}}}$ or ${Q _{\rm{max}}}$ and the corresponding feedback performance as the final output.
}

\begin{algorithm}[t]
        \caption{BNA algorithm}
        \label{algorithm1}
        \KwIn{Initial feedback bit number ${{Q_{\rm{i}}}}$ and corresponding SGCS performance ${{\rho_{\rm{i}}}}$; Target SGCS performance ${{\rho_{\rm{t}}}}$; Supported minimum feedback bit number ${{Q_{\rm{min}}}}$ and maximum feedback bit number ${{Q_{\rm{max}}}}$; Tolerance Error $\varepsilon = 0.01$}
        \KwOut{Adjusted feedback bit number ${{Q_{\rm{a}}}}$ and corresponding SGCS performance ${{\rho_{\rm{a}}}}$}
        \BlankLine
        \uIf{$\left| {{\rho _{\rm{i}}} - {\rho _{\rm{t}}}} \right| \le \varepsilon$}
        {
            $\left[ {{Q_{\rm{a}}},{\rho _{\rm{a}}}} \right] = \left[ {{Q_{\rm{i}}},{\rho _{\rm{i}}}} \right]$;
        }
        \uElseIf{${\rho _{\rm{i}}} - {\rho _{\rm{t}}} > \varepsilon$}
        {
            $\left[ {{Q_{\rm{a}}},{\rho _{\rm{a}}}} \right] = $ \textit{Binary\_Search(${Q_{\rm{min}}}$, ${Q_{\rm{i}}}$)};
        }
        \ElseIf{${\rho _{\rm{i}}} - {\rho _{\rm{t}}} < -\varepsilon$}
        {
            $\left[ {{Q_{\rm{a}}},{\rho _{\rm{a}}}} \right] = $ \textit{Binary\_Search(${Q_{\rm{i}}}$, ${Q_{\rm{max}}}$)};
        }
    \Return{$Q_{\rm{a}}, \rho _{\rm{a}}$}; \par
    \BlankLine
    \SetKwProg{Fn}{Function}{:}{end} 
    \Fn{Binary\_Search($Q_{\rm{l}}$, $Q_{\rm{r}}$)}
    {
        \While{${Q_{\rm{l}}} \le {Q_{\rm{r}}}$} 
            { 
                ${Q_{\rm{mid}}} = {{\left( {{Q_{\rm{l}}} + {Q_{\rm{r}}}} \right)} \mathord{\left/
 {\vphantom {{\left( {{Q_{\rm{l}}} + {Q_{\rm{r}}}} \right)} 2}} \right.
 \kern-\nulldelimiterspace} 2}$; \par
                Calculate the SGCS $\rho_{\rm{mid}}$ as (\ref{eq9}) with $Q_{\rm{mid}}$; \par
                \uIf{$\left| {{\rho _{\rm{mid}}} - {\rho _{\rm{t}}}} \right| \le \varepsilon$} 
                { 
                    \Return{${Q_{\rm{mid}}}, \rho_{\rm{mid}}$};
                }
                \uElseIf{${\rho _{\rm{mid}}} - {\rho _{\rm{t}}} < -\varepsilon$}
                {
                    $Q_{\rm{l}} = {Q_{\rm{mid}}} + 1$;
                }
                \Else
                {
                    $Q_{\rm{r}} = {Q_{\rm{mid}}} - 1$;
                }
            }
            \Return{${Q_{\rm{mid}}}, \rho_{\rm{mid}}$};
    }
\end{algorithm}


\subsection{UE-First Separate Training for Different Manufacturers}

\subsubsection{Training process} When the UE and gNB come from different manufacturers, the trained encoder that matched to the adopted decoder at the gNB may not be transmitted to the UE due to the model protection. {\color{black}On the other hand, most existing studies only consider the CSI feedback between a single UE and a single gNB.} However, when multiple UEs perform CSI feedback simultaneously, the encoder structure of each UE may be different due to the variable requirements for feedback performance. {\color{black}Hence, the gNB needs to store
the decoder corresponding to each UE’s encoder for CSI reconstruction, which will consume the gNB storage resources and make the system inefficient \cite{r6_new, r11_new}.}

Therefore, as illustrated in Fig. 6, a novel separate training approach with multiple UEs and a single gNB is proposed. It is assumed that the UE is not limited to the mobile phone and has sufficient ability for AI training with the technology development \cite{r9_new, r10_new}. The main steps of separate training are given as follows:
\begin{itemize}[leftmargin=*]
	\item[$\bullet$] The $n$th UE$\left( {n = 1,2, \cdots ,N} \right)$ first trains its own CSI feedback model with the data set $\left\{ {{{\bf{w}}_n}} \right\}$. After training, the $n$th UE only employs its own encoder network, denoted as ${\rm{Encoder}}\ n$.
\end{itemize}
\begin{itemize}[leftmargin=*]
	\item[$\bullet$] Then, the $n$th UE transmits its own input data set $\left\{ {{{\bf{w}}_n}} \right\}$ and the corresponding output feedback bit streams $\left\{ {{{\bf{s}}_n}} \right\}$ of ${\rm{Encoder}}\ n$ to the gNB through the uplink.
\end{itemize}
\begin{itemize}[leftmargin=*]
	\item[$\bullet$] Finally, the gNB utilizes all the data received from all UEs, denoted as $\left\{ {{{\bf{w}}_1},{{\bf{w}}_2}, \cdots {{\bf{w}}_N};{{\bf{s}}_1},{{\bf{s}}_2}, \cdots {{\bf{s}}_N}} \right\}$,  to train a general decoder to reconstruct CSI eigenvector of different UEs.
\end{itemize}
\begin{figure}[t]
	\centering
	\includegraphics[width=\mysinglefigwidth]{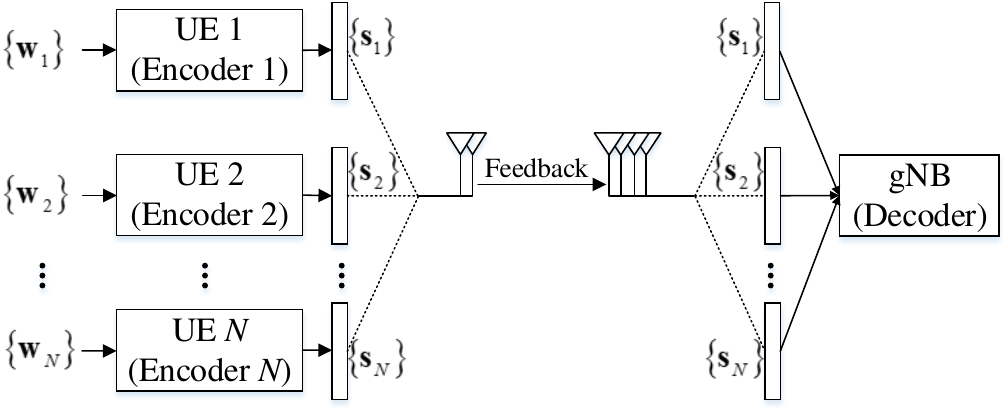}
	\caption{CSI feedback with multiple UEs and a single gNB.}
	\label{FIGURE5}
	\vspace{-0.2cm} 
\end{figure}

Thus, the model protection problem between the UEs and gNB from different manufacturers is resolved by the proposed separate training approach. At the same time, a single decoder for CSI eigenvector recovery at the gNB can save the gNB's storage capacity.

\vspace{0.1cm}
\subsubsection{Loss function} For separate training, SGCS can still be used as the loss function for the training of each UE’s feedback model and the training of decoder at the gNB. Specifically, the loss function ${L_2}$ of training decoder at the gNB can be written as
\begin{flalign}\label{15}
	{L_2} = \frac{1}{{NK}}{\sum\limits_{n = 1}^N {\sum\limits_{k = 1}^K {\left( {\frac{{\left\| {{\bf{w}}_{nk}^{\rm{H}}{{{\bf{\hat w}}}_{nk}}} \right\|}}{{\left\| {{{\bf{w}}_{nk}}} \right\|\left\| {{{{\bf{\hat w}}}_{nk}}} \right\|}}} \right)} } ^2},
\end{flalign}
where ${{\bf{w}}_{nk}}$ represents the input eigenvector of the \textit{k}th subband of the \textit{n}th UE and ${{\bf{\hat w}}_{nk}}$ is the recovered eigenvector corresponding to ${{\bf{w}}_{nk}}$.


\vspace{0.1cm}
\subsubsection{Practicability of separate training} Because the proposed separate training approach has no limitation on model structure, it can be applied to any kind of traditional CSI feedback models, {\color{black}including the proposed ABLNet with FBCU.} Moreover, the same encoder structure or different structure for multiple UEs can be adopted during the training phase. In this way, each UE’s encoder can compress CSI eigenvector with different subband numbers and quantize the codeword into length-variable feedback bit stream by using FBCU, and then the single decoder at the gNB is used to dequantize and reconstruct the CSI eigenvector corresponding to each UE.

\section{Simulation Results and Analyses}
\label{section4}

In this section, the dataset, the relevant training setup and the structure of networks are first introduced in detail. Then, a series of experiments are implemented to evaluate the performance of the proposed ABLNet for the CSI eigenvectors with different numbers of subbands. Furthermore, the performance of ABLNet with FBCU is discussed for length-variable feedback bit streams. {\color{black}After that, the adjustment effect of the BNA algorithm is illustrated. And the link-level block error rate (BLER) performance with different feedback schemes is analyzed.} Finally, when different UEs have the same or different encoder structure, the performance of separate training with multiple UEs and a single gNB is evaluated. 

\begin{table}[t]
	\caption{\rmfamily{ABLNet model structure.}}\label{Table1}
	\centering
	\renewcommand{\arraystretch}{1.8}
	\begin{tabular}{|c|c|lll}
		\cline{1-2}
		Model                    & Structure                    &  &  &  \\ \cline{1-2}
		\multirow{6}{*}{Encoder} & Input layer, ${\rm{input}} = \left( {{K_{\max }},64} \right)$              &  &  &  \\ \cline{2-2}
		& BiLSTM, ${\rm{input}} = \left( {{K_{\max }},64} \right)$,  ${\rm{output}} = \left( {{K_{\max }},256} \right)$              &  &  &  \\ \cline{2-2}
		& BiLSTM, ${\rm{input}} = \left( {{K_{\max }},256} \right)$, ${\rm{output}} = \left( {{K_{\max }},1024} \right)$              &  &  &  \\ \cline{2-2}
		& Dense1, ${\rm{input}} = \left( {{K_{\max }},1024} \right)$, ${\rm{output}} = \left( {{K_{\max }},64} \right)$             &  &  &  \\ \cline{2-2}
		& Dense2, ${\rm{input}} = \left( {{K_{\max }},64} \right)$, ${\rm{output}} = \left( {{K_{\max }},5} \right)$   &  &  &  \\ \cline{2-2}
		& Quantization layer, ${\rm{input}} = K \times 5$, ${\rm{output}} = K \times 10$   &  &  &  \\ \cline{1-2}
		\multirow{6}{*}{Decoder}& Dequantization layer, ${\rm{input}} = K \times 10$, ${\rm{output}} = K \times 5$ &  &  &  \\ \cline{2-2}
		& Dense3, ${\rm{input}} = \left( {{K_{\max }},5} \right)$, ${\rm{output}} = \left( {{K_{\max }},64} \right)$             &  &  &  \\ \cline{2-2}
		& BiLSTM, ${\rm{input}} = \left( {{K_{\max }},64} \right)$, ${\rm{output}} = \left( {{K_{\max }},256} \right)$              &  &  &  \\ \cline{2-2}
		& BiLSTM, ${\rm{input}} = \left( {{K_{\max }},256} \right)$, ${\rm{output}} = \left( {{K_{\max }},1024} \right)$              &  &  &  \\ \cline{2-2}
		& Dense1, ${\rm{input}} = \left( {{K_{\max }},1024} \right)$, ${\rm{output}} = \left( {{K_{\max }},64} \right)$             &  &  &  \\ \cline{2-2}
		& Output layer, ${\rm{output}} = \left( {{K},64} \right)$             &  &  &  \\ \cline{1-2}
	\end{tabular}
	\vspace{-0.1cm}
\end{table}

\subsection{Simulation Configuration}
{\color{black}{The clustered delay line (CDL) channel model defined in the 3GPP is adapted to generate data samples and two types of channels with delay spread 30ns are taken into consideration: CDLA and CDLC.}} Each channel contains three kinds of subband number for CSI eigenvector, which are $K=3$ subbands, $K=6$ subbands and $K=12$ subbands. Moreover, a typical setup that ${N_{\rm{T}}}=32$ transmit antennas at the gNB and ${N_{\rm{R}}}=4$ receive antennas at UE is adopted in the MIMO-OFDM system. Therefore, the size of CSI eigenvector is $K \times 2{N_{\rm{T}}} = K \times 64\left( {K = 3,6,12} \right)$. 

At the same time, the sizes of training and testing datasets are 50,000 and 1,000 respectively for CSI eigenvector with any subband number. For the training phase, we set the numbers of training epochs and batch size as 500 and 128, respectively. The initial learning rate is 0.0005, which will be adaptively adjusted with the training process. The default adaptive momentum (Adam) optimizer is employed. Because the proposed FBCU is a plug-in module and compatible with the proposed network, so the training setup is the same as the original setup mentioned above when training models with FBCU. Additionally, we fix the quantization bit number $q = 2$.

The detailed model structure of ABLNet is indicated in Table \ref{Table1}. Because the input layer of the model is fixed, CSI eigenvectors with different subband numbers should be unified into the same dimension ${K_{\max }} \times 64$ by padding zero according to the maximum subband number ${K_{\max }}$. 

Furthermore, to verify the SGCS performance of the ABLNet and FBCU with different subband numebrs of input CSI eigenvectors, the designed model structure is utilized to train CSI eigenvectors with a fixed subband number, denoted as BiLSTMNet. And the obtained SGCS performance of BiLSTMNet is used as the benchmark.


\subsection{Performance of Proposed ABLNet}
\label{4.2}
The number of feedback bits is assumed to be $Q=120$ bits for CSI eigenvector with $K=12$ subbands, and then the corresponding number of CSI feedback bits for $K=3$ subbands eigenvector and $K=6$ subbands eigenvector are $Q=30$ bits and $Q=60$ bits, respectively. 


\begin{figure}[!t]
	\centering
	\includegraphics[width=\mysinglefigwidth]{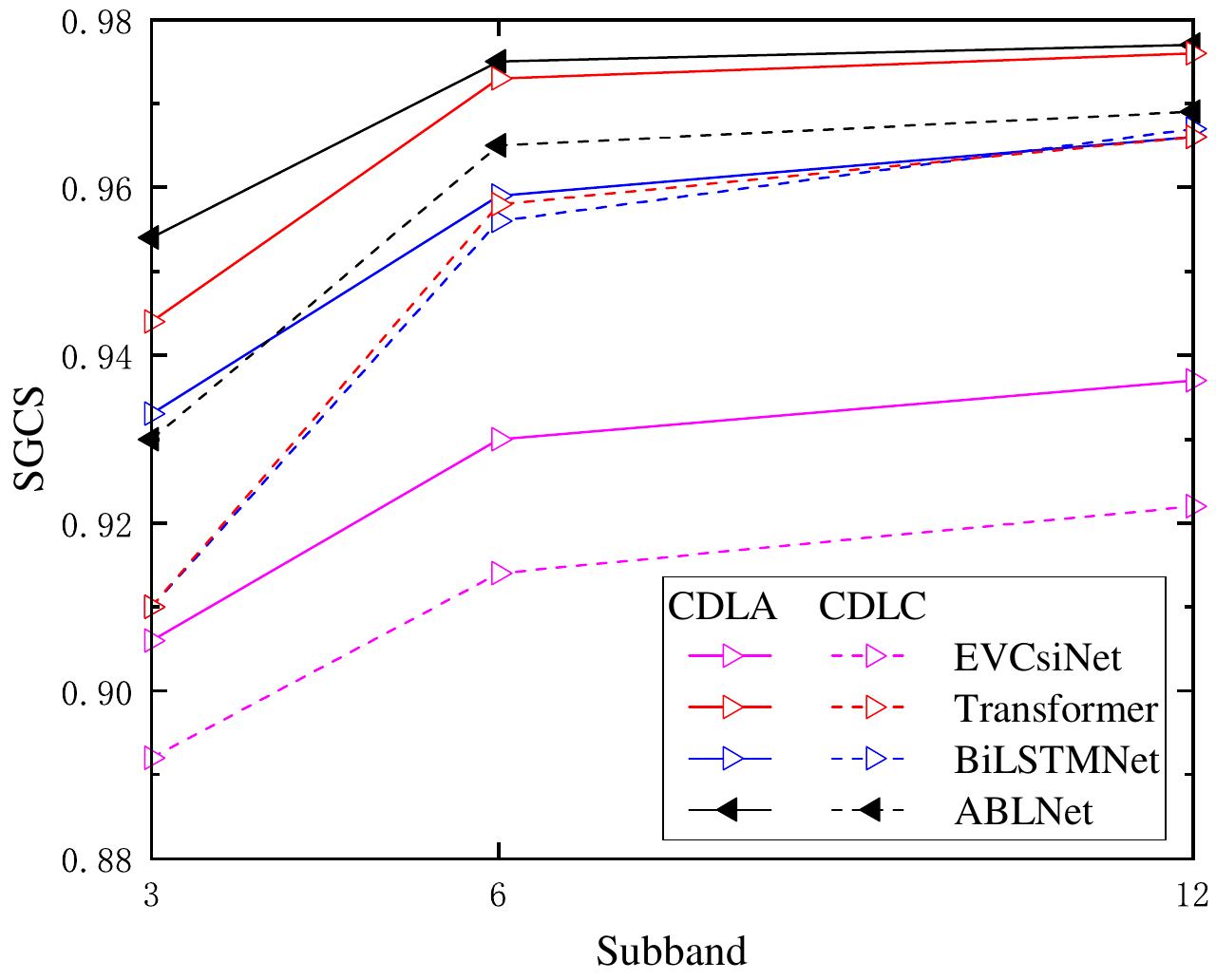}
	\caption{Feedback performance with different subband number.}
	\label{FIGURE7}
	\vspace{-0.2cm} 
\end{figure}

{\color{black}To verify the feedback performance of the ABLNet model, different comparison models, such as EVCsiNet and Transformer \cite{refer27, refer28}, for CSI eigenvector with different subband numbers are trained under both CDLA and CDLC channel, as illustrated in Fig. \ref{FIGURE7}. Concretely, the adaptive input scheme in \cite{add4} is adopted to train both EVCsiNet and Transformer. In Fig. \ref{FIGURE7}, it can be seen that the feedback performance of ABLNet is better than that of BiLSTMNet and EVCsiNet under $K=3$ subbands and $K=6$ subbands and similar to the Transformer under $K=12$ subbands for CDLA channel. Similarly, the feedback performance of ABLNet for CDLC channel outperforms those of other models under any kind of subband number.}
\begin{table}[!t]
\caption{\rmfamily{\color{black}Model complexity comparison.}}\label{Table2}
\centering
\setlength{\tabcolsep}{2.6mm}
\renewcommand{\arraystretch}{2.1}
\begin{tabular}{|c|c|c|}
\hline
Feedback Scheme & Trainable Parameters ($\times {10^6}$) & FLOPs ($\times {10^7}$) \\ \hline
EVCsiNet        & 6.98                                        & 1.41                         \\ \hline
Transformer     & 23.95                                       & 51.29                        \\ \hline
ABLNet          & 14.75                                       & 32.40                        \\ \hline
\end{tabular}
\end{table}

Moreover, the SGCS gain of low subband number is more obvious. Because compared with the BiLSTMNet trained by a fixed subband number alone, the ABLNet trained with multiple subband numbers can learn more data features of other subband. Additionally, with the increase of subband number for each curve in Fig. \ref{FIGURE7}, the feedback performance also increases gradually, because more subbands make CSI eigenvector contain more features that can be extracted during training. 

{\color{black}Additionally, as listed in Table \ref{Table2}, the model complexity of different feedback schemes are analyzed in terms of trainable parameters and floating point operations (FLOPs). We can see that the proposed ABLNet has fewer trainable parameters and FLOPs than Transformer, but has better feedback performance. Although EVCsiNet has the fewest trainable parameters and FLOPs, its performance is worst among these feedback schemes.}

Therefore, simulation results show that the proposed ABLNet can adapt to the CSI eigenvector with different subband numbers {\color{black}and has a low model complexity while high CSI recovery accuracy}; moreover, the feedback bit number can be proportionally varied with the number of subbands.
\vspace{-0.2cm}
\subsection{Performance of ABLNet with FBCU}

\begin{figure}[t]
	\vspace{0.1cm} 
	\centering
	\includegraphics[width=\mysinglefigwidth]{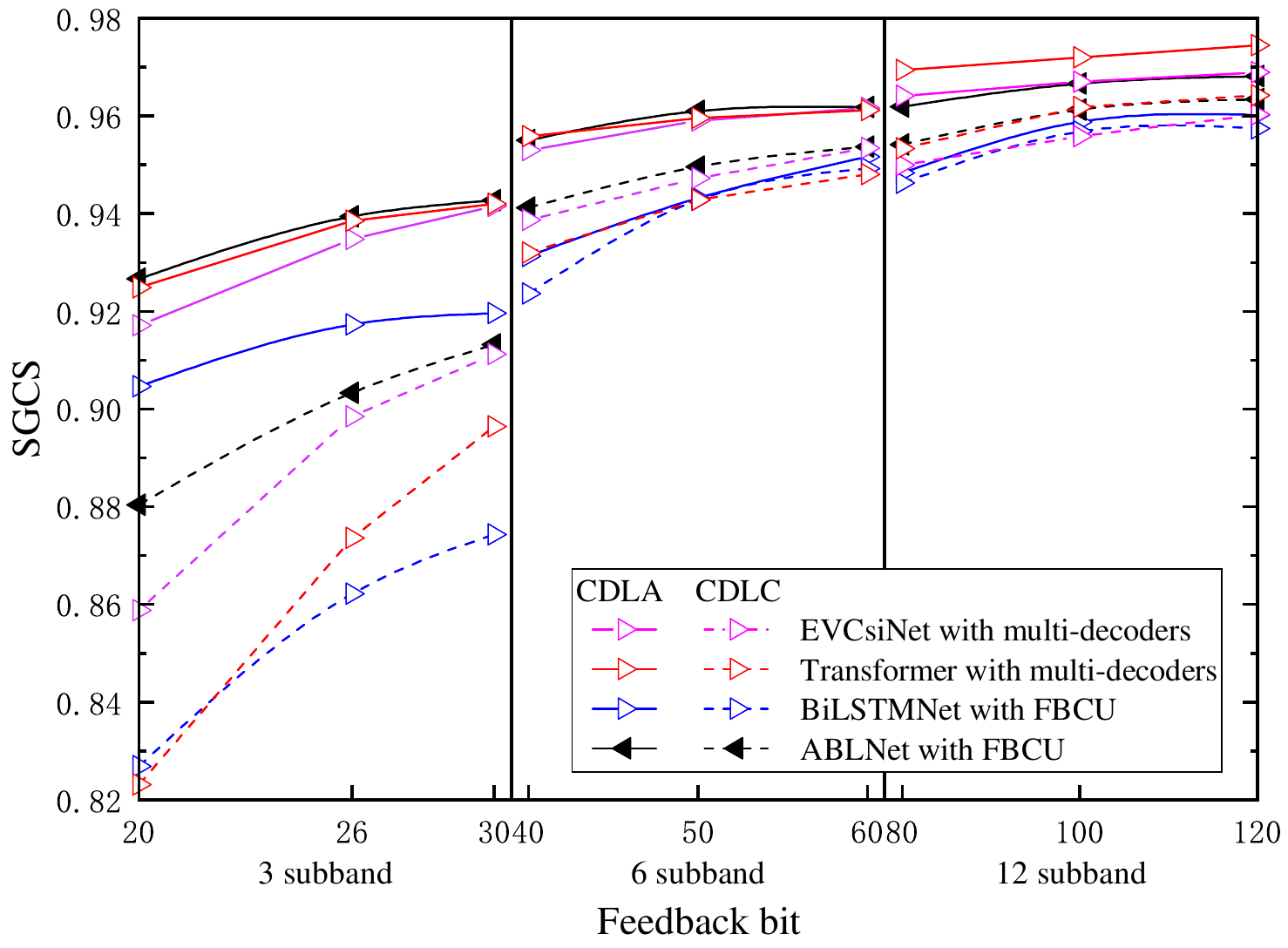}
	\caption{Feedback performance with different subband numbers and feedback bits.}
	\label{add}
	\vspace{-0.1cm} 
\end{figure}

{\color{black}The performance of ABLNet with FBCU is tested under the feedback bit number $Q=20$ to $30$ for $K=3$ subbands, $Q=40$ to $60$ for $K=6$ subbands and $Q=80$ to $120$ for $K=12$ subbands, respectively.} 
{\color{black}Fig. \ref{add} illustrates the SGCS curves under both the ABLNet with FBCU and BiLSTMNet with FBCU. Moreover, the SGCS performance of EVCsiNet and Transformer with single encoder and multiple decoders is taken as the benchmark \cite{refer27, refer28, refer32}. To save the training cost, three decoders for EVCsiNet and Transformer are employed, as well as $Q = \left\{ {20,26,30} \right\}$ for $K=3$ subbands, $Q = \left\{ {40,50,60} \right\}$ for $K=6$ subbands and $Q = \left\{ {80,100,120} \right\}$ for $K=12$ subbands, respectively. Whether for CDLA or CDLC channel, it is obvious that the proposed ABLNet trained with FBCU outperforms other training approaches under $K=3$ subbands and $K=6$ subbands. While for $K=12$ subbands, the SGCS performance of ABLNet with FBCU is slightly less than that of EVCsiNet and Transformer with multi-decoders. The reason has been mentioned above that data features of eigenvectors with small subband number have a negative impact on the eigenvectors with large subband number during the training process.}
%
		
At the same time, the SGCS performance is gradually improved with the increase of feedback bit number for each subband number for each training approach, because the longer the feedback bit stream is, the more feature information it contains and this will help the decoder to reconstruct the CSI eigenvectors.




{\color{black}In summary, simulation results demonstrate that the FBCU module is compatible with the proposed ABLNet model and makes the model adapt to different subband numbers of CSI eigenvectors and different feedback bit numbers at the same time.}
{\color{black}\subsection{Performance of BNA Algorithm}

To verify the adjustment effect of BNA algorithm, the ABLNet with FBCU is first trained under $Q=20$ to $120$ for $K=12$ subbands. Then, $Q=49$ of enhanced TypeII (eTypeII) method is taken as the benchmark, and the corresponding target SGCS performance are ${\rho _{\rm{t}}}=0.904$ and ${{\rho_{\rm{t}}}}=0.840$ for CDLA and CDLC channel, respectively \cite{refer27}. Additionally, the initial feedback bit number ${Q _{\rm{i}}}=48$ is chosen for ABLNet with FBCU. The other subband numbers have the similar conclusions and therefore are ignored here.

\begin{figure}[t]
	\centering
	\includegraphics[width=\mysinglefigwidth]{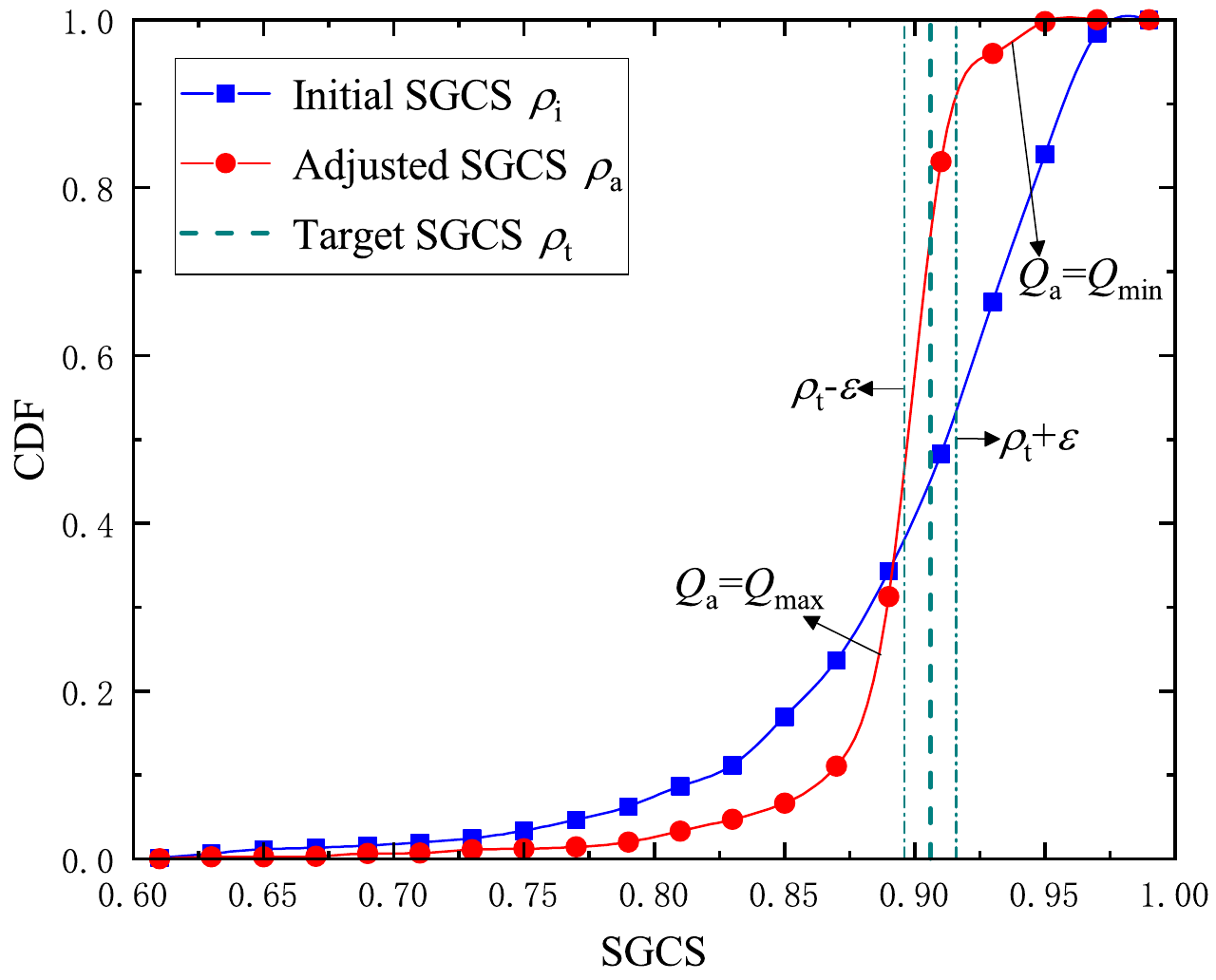}
	\caption{{\color{black}Adjustment effect of BNA algorithm for CDLA.}}
	\label{FIGURE_adjust_CDLA}
\end{figure}

\begin{figure}[t]
	\centering
	\includegraphics[width=\mysinglefigwidth]{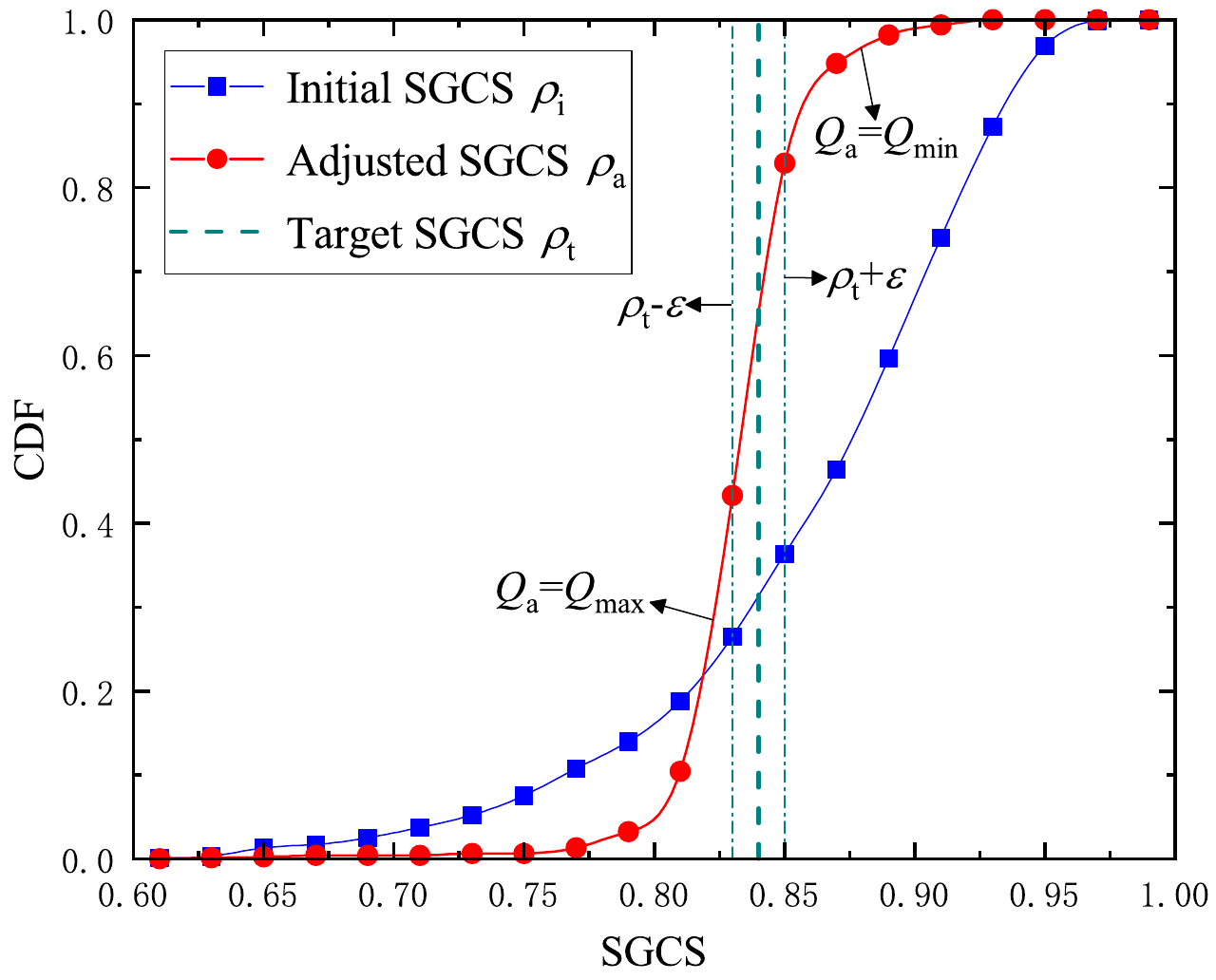}
	\caption{{\color{black}Adjustment effect of BNA algorithm for CDLC.}}
	\label{FIGURE_adjust_CDLC}
	\vspace{-0.2cm} 
\end{figure}

The cumulative distribution function (CDF) is taken to illustrate the SGCS performance changes of input CSI before and after the adaptive adjustment, as ${{\rho_{\rm{i}}}}$ and ${{\rho_{\rm{a}}}}$ illustrated in both Fig. \ref{FIGURE_adjust_CDLA} and Fig. \ref{FIGURE_adjust_CDLC}. From both figures, it can be obviously noticed that the CDF of ${{\rho_{\rm{a}}}}$ are higher than that of ${{\rho_{\rm{i}}}}$ within the tolerance error margin of ${{\rho_{\rm{t}}}}$, meaning that the SGCS performance of most input CSI are effectively adjusted close to the target performance ${{\rho_{\rm{t}}}}$ through the BNA algorithm. Specifically, when the SGCS ${{\rho_{\rm{a}}}}$ reaches ${\rho _{\rm{t}}} + \varepsilon $, the CDF improves about 40\% and 50\% compared with ${{\rho_{\rm{i}}}}$ for CDLA and CDLC, respectively. Moreover, the feedback bit number of input CSI eigenvectors with SGCS lower than ${\rho _{\rm{t}}} - \varepsilon $ has been adjusted as ${Q _{\rm{a}}}={Q _{\rm{max}}}$, which has improved the SGCS performance of every input CSI as much as possible; and for the input CSI eigenvectors with SGCS higher than ${\rho _{\rm{t}}} + \varepsilon $, the feedback bit number has been adjusted to ${Q _{\rm{a}}}={Q _{\rm{min}}}$ for saving the average feedback overhead. Additionally, the average value of feedback bit number ${{Q_{\rm{a}}}}$ is also calculated, which is 44 bits for both two channels and save 4 bits compared with ${{Q_{\rm{i}}}}$.

Therefore, the results demonstrate that the BNA algorithm can adaptively adjust the feedback bit number of every input CSI to make the SGCS performance close to the target SGCS performance ${{\rho_{\rm{t}}}}$ and can also save the average feedback bit number ${{Q_{\rm{a}}}}$ after adjustment.
}

\begin{figure}[t]
	\centering
	\includegraphics[width=\mysinglefigwidth]{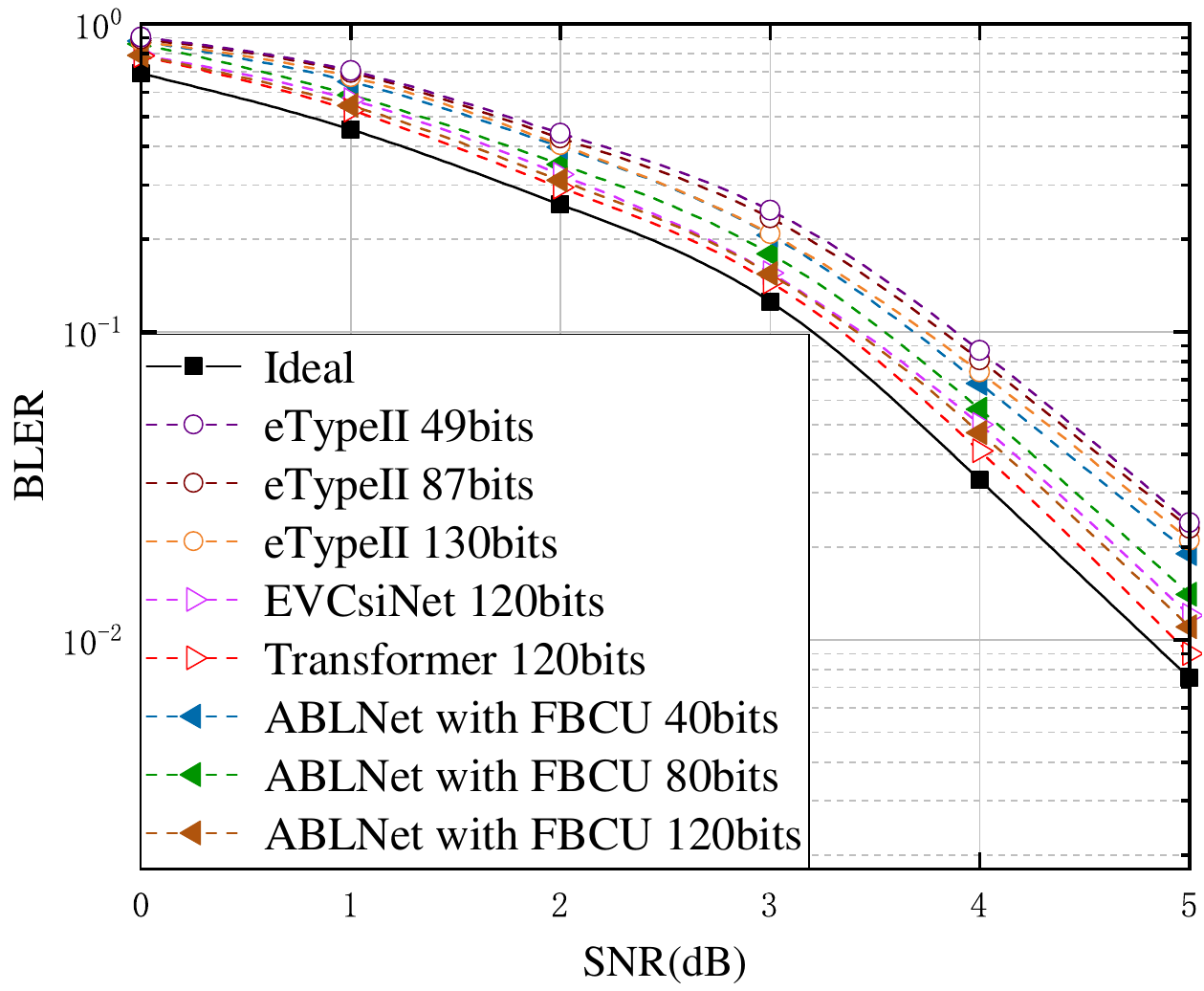}
	\caption{\color{black}Link-level BLER performance comparison for different CSI feedback schemes.}
	\label{BLER}
	\vspace{-0.2cm} 
\end{figure}

{\color{black}\subsection{Performance of Link-Level BLER}

As shown in Fig. \ref{BLER}, the link-level BLER for different CSI feedback schemes is depicted under CDLA channel. The ideal condition means that the gNB knows the CSI eigenvector and therefore the SGCS $\rho=1$. Meanwhile, the BLER performance of $Q = \left\{ {40,80,120} \right\}$ for ABLNet with FBCU is compared to those of EVCsiNet and Transformer at $Q = 120$ bits. {\color{black}Moreover, for the traditional CSI feedback scheme, i.e., eTypeII, the number $Q = \left\{ {49,87,130} \right\}$ of feedback bits is close to those of DL-based CSI feedback schemes and therefore is adopted here for comparison.}


It can be seen that the proposed ABLNet with FBCU achieves about ${\rm{BLER}} = {10^{ - 2}}$ when $Q = 120$ and ${\rm{SNR}} = {5}$, which significantly outperforms eTypeII with $Q = 130$ bits; the proposed ABLNet with FBCU and $Q = 40$ bits can achieve similar BLER performance to eTypeII with $Q = 130$ bits. As the feedback bit number $Q$ increases for both the ABLNet with FBCU and eTypeII, the BLER performance decreases gradually; because the gNB can select a more appropriate beamforming vector with larger SGCS performance for each input CSI eigenvector, which is more beneficial to the average BLER reduction. Moreover, it can be noticed that three different DL-based CSI feedback models have similar BLER performance with $Q = 120$ feedback bits.

Therefore, the results show the superiority of the proposed ABLNet with FBCU on the BLER performance and feedback overhead compared with conventional eTypeII method. Additionally, the proposed ABLNet with FBCU can obtain the similar BLER performance of other DL-based CSI feedback schemes with the capability of adaptive input length of CSI eigenvectors and output number of feedback bits.
}

\subsection{Performance of UE-First Separate Training}

This subsection mainly verifies the recovery accuracy of CSI eigenvectors by using a general decoder for multiple UEs (where the multiple encoders and a general decoder are separately trained) compared with multiple separate decoders (where the encoder and decoder for each UE are jointly trained). Assume that three UEs simultaneously perform CSI feedback with the subband number $K=12$ and the corresponding feedback bit number $Q=120$ for CSI eigenvector reconstruction. The other subband numbers have the same conclusions and therefore are ignored here for saving the space. 

\begin{figure}[t]
	\centering
	\includegraphics[width=\mysinglefigwidth]{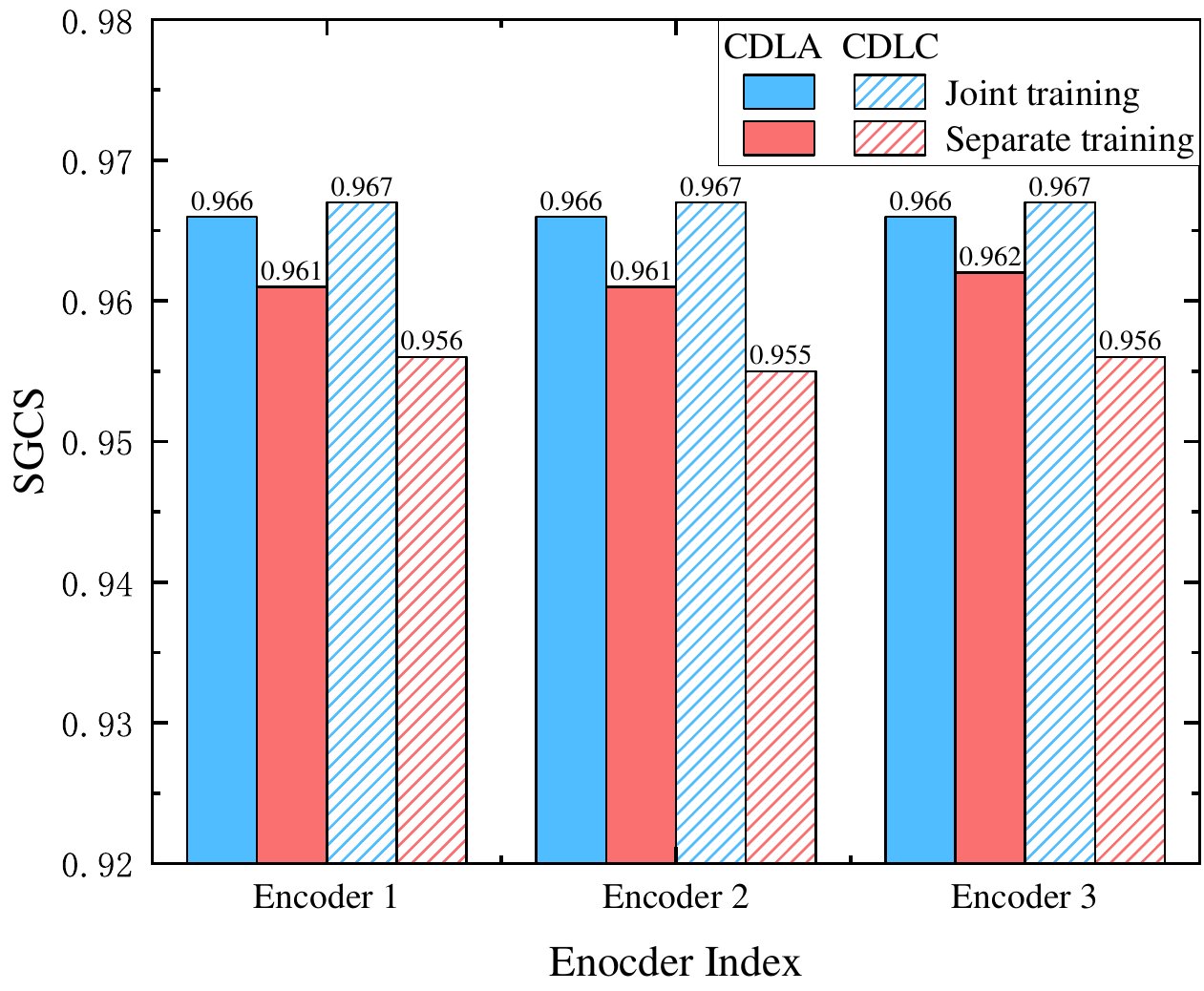}
	\caption{Feedback performance of same encoders under same CSI distribution.}
	\label{FIGURE10}
\end{figure}

\begin{table}[!t]
\caption{\rmfamily{Encoder structure of different UEs.}}\label{Table3}
	\rmfamily
        \centering
	\setlength{\tabcolsep}{1.8mm}
	\renewcommand{\arraystretch}{1.6}
\begin{tabular}{|c|c|}
\hline
Model                     & Structure                                       \\ \hline
Encoder 1                  & {\color{black}The same as the structure of Encoder in Table I.} \\ \hline
\multirow{6}{*}{Encoder 2} & Input layer, ${\rm{input}} = \left( {{K_{\max }},64} \right)$           \\ \cline{2-2} 
		& BiLSTM, ${\rm{input}} = \left( {{K_{\max }},64} \right)$, ${\rm{output}} = \left( {{K_{\max }},512} \right)$            \\ \cline{2-2} 
		& BiLSTM, ${\rm{input}} = \left( {{K_{\max }},512} \right)$, ${\rm{output}} = \left( {{K_{\max }},2048} \right)$            \\ \cline{2-2} 
		& Dense1, ${\rm{input}} = \left( {{K_{\max }},2048} \right)$, ${\rm{output}} = \left( {{K_{\max }},64} \right)$           \\ \cline{2-2} 
		& Dense2, ${\rm{input}} = \left( {{K_{\max }},64} \right)$, ${\rm{output}} = \left( {{K_{\max }},5} \right)$          \\ \cline{2-2} 
		& Quantization layer, ${\rm{input}} = K \times 5$, ${\rm{output}} = K \times 10$ \\ \hline
\multirow{6}{*}{Encoder 3} & Input layer, ${\rm{input}} = \left( {{K_{\max }},64} \right)$            \\ \cline{2-2} 
		& GRU, ${\rm{input}} = \left( {{K_{\max }},64} \right)$, ${\rm{output}} = \left( {{K_{\max }},256} \right)$                \\ \cline{2-2} 
		& GRU, ${\rm{input}} = \left( {{K_{\max }},256} \right)$, ${\rm{output}} = \left( {{K_{\max }},1024} \right)$                \\ \cline{2-2} 
		& Dense1, ${\rm{input}} = \left( {{K_{\max }},1024} \right)$, ${\rm{output}} = \left( {{K_{\max }},64} \right)$           \\ \cline{2-2} 
		& Dense2, ${\rm{input}} = \left( {{K_{\max }},64} \right)$, ${\rm{output}} = \left( {{K_{\max }},5} \right)$           \\ \cline{2-2} 
		& Quantization layer, ${\rm{input}} = K \times 5$, ${\rm{output}} = K \times 10$ \\ \hline
\end{tabular}
\vspace{-0.1cm}
\end{table}

\subsubsection{Same encoder structure of UEs} When each UE has the same encoder structure, the encoder and decoder structure of the proposed ABLNet are given in Table \ref{Table1}. The comparison results between the separate training and joint training for the three UEs are illustrated in Fig. \ref{FIGURE10}. We can see that the feedback performance under the separate training is slightly lower than that under the joint training for each UE. Specifically, for CDLA channel, the SGCS of general decoder under the separate training only decreases by 0.0046 on average compared to that of separate decoder under the joint training; however, the SGCS decreases by 0.0113 on average for CDLC channel.


		
		
\subsubsection{Different encoder structure of UEs} {\color{black}When each UE takes different encoder structure to meet actual feedback requirements, we mainly change the encoder structure by modifying the number of LSTM cell neurons or replacing LSTM cells, which is illustrated in detail in Table \ref{Table3}. Encoder 1 directly takes the encoder structure of the proposed ABLNet, Encoder 2 increases the input and output dimension of LSTM cell, and Encoder 3 replaces LSTM cell with GRU \cite{refer36} to simplify the network complexity. In addition, the general decoder takes the structure of ABLNet decoder which is shown in Table \ref{Table1}.}

\begin{figure}[t]
	\centering
	\includegraphics[width=\mysinglefigwidth]{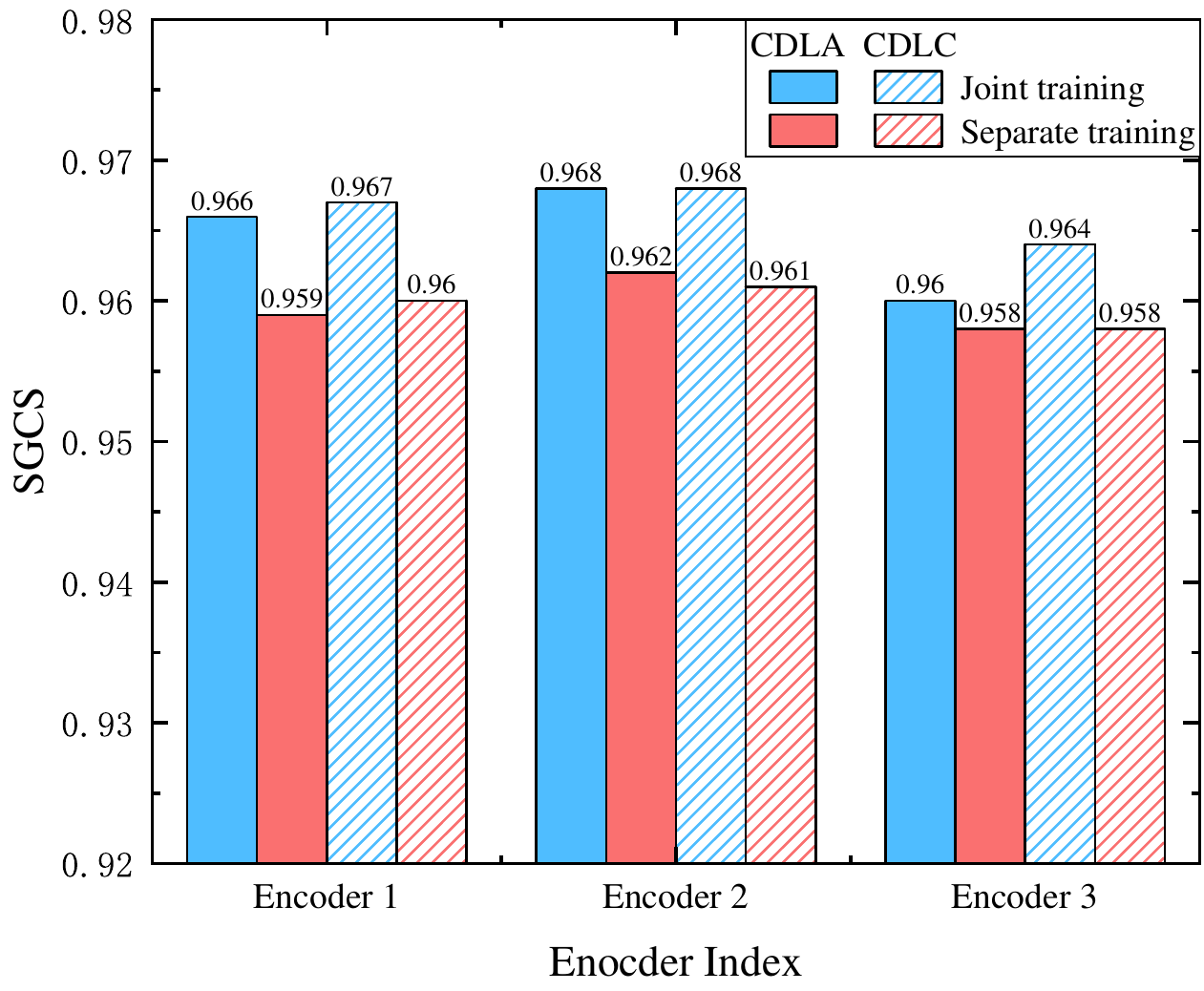}
	\caption{Feedback performance of different encoders under same CSI distribution.}
	\label{FIGURE11}
	\vspace{-0.2cm} 
\end{figure}

From Fig. \ref{FIGURE11}, it can be noticed that the separate training approach with general decoder has a similar SGCS performance to that of joint training for each UE with separate decoder. Specifically, the feedback performance degrades by 0.007, 0.006 and 0.002 for Encoder 1, Encoder 2 and Encoder 3 under CDLA channel; and degrades by 0.007, 0.007 and 0.006 under CDLC channel.

\begin{figure}[t]
	\centering
	\includegraphics[width=\mysinglefigwidth]{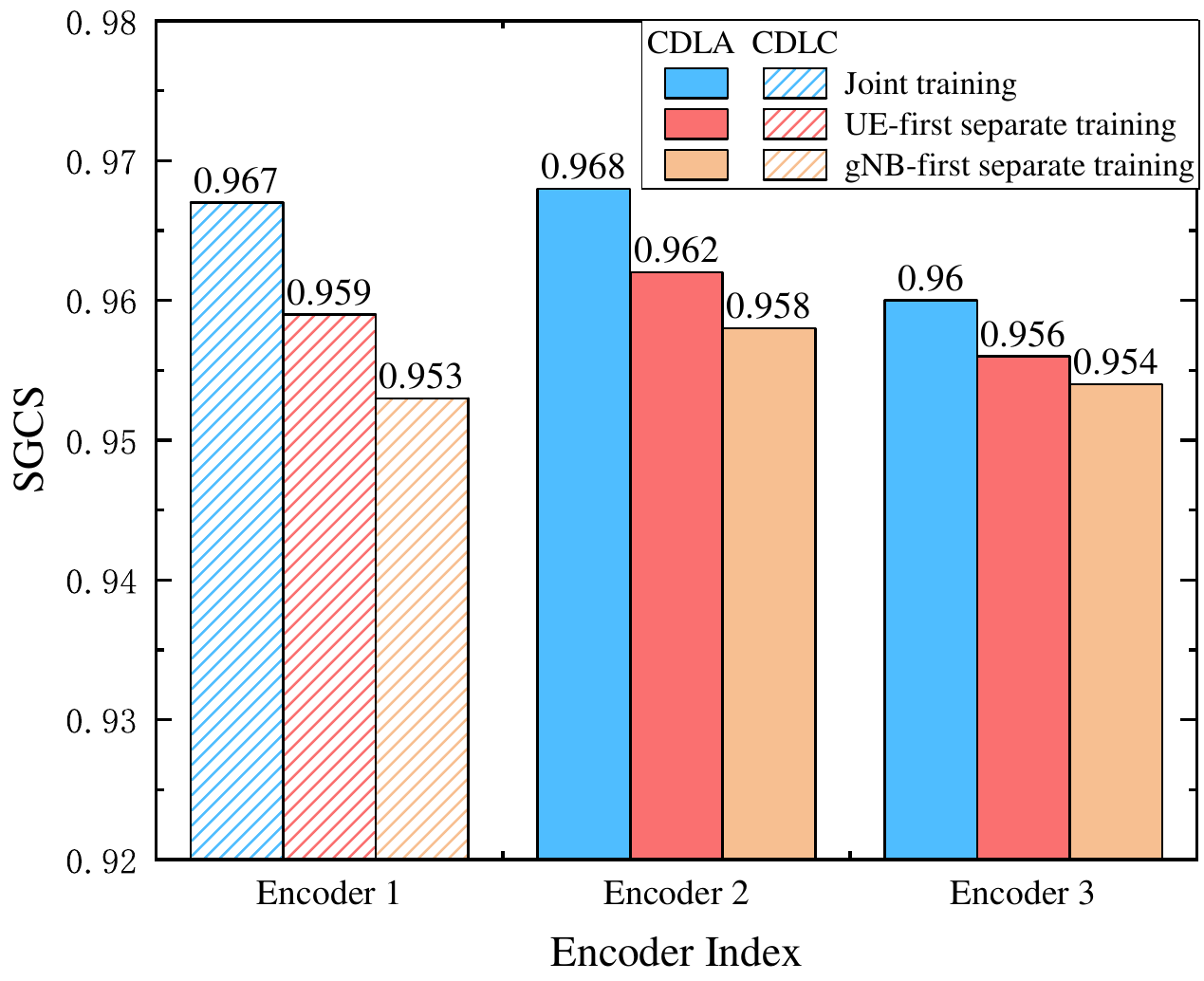}
	\caption{\color{black}{Feedback performance of different encoders under different CSI distribution.}}
	\label{FIGURE_new}
	\vspace{-0.4cm} 
\end{figure}

		
		

Additionally, Encoder 2 has the best feedback performance, followed by Encoder 1 and Encoder 3. The main reason is that Encoder 2 has large number of neurons and can enhance the learning ability and extract more data features. On the other hand, the encoder with GRU has different structure with the decoder, and thus results in weaker representation ability of the model than the original model structure.

Moreover, to verify the feasibility of separate training under different CSI distributions, Encoder 1 under CDLC channel as well as both Encoder 2 and Encoder 3 under CDLA channel are studied for the two training approaches. The results in Fig. \ref{FIGURE_new} indicate that even under different channels, the SGCS performance of UE-first separate training is still close to that of joint training, which is consistent with the results trained under the same distributed CSI. 

{\color{black}At the same time, the performance of gNB-first separate training strategy for multi-UEs feedback is evaluated against the UE-first separate training strategy. To ensure a fair comparison, the gNB first trains the ABLNet under both CDLA and CDLC channels; then, the Encoder 1 is trained under CDLC channel while the Encoder 2 and Encoder 3 are trained under CDLA channel; finally, the input and output of all encoders are utilized to fine-tune the decoder of gNB to reduce the performance loss as \cite{r15_new} did. According to Fig. \ref{FIGURE_new}, the UE-first separate training performs better than the gNB-first separate training, which further reveals the feasibility of the proposed UE-first separate training strategy for multi-UEs feedback.}

Therefore, simulation results show that training multiple encoders at UE and a general decoder at the gNB by separate training approach can achieve similar performance to joint training for each UE, {\color{black}even for different CSI distributions}.

\section{Conclusions}
\label{section5}

This paper proposes an adaptive CSI feedback model ABLNet and designs a FBCU to process different lengths of input CSI eigenvectors and numbers of feedback bits. {\color{black}Then, a BNA algorithm is developed to achieve a target SGCS for every input CSI by adjusting feedback bit number flexibly.} Moreover, a UE-first separate training approach is realized for decoupled training of encoders and decoder among the UEs and gNB from different manufacturers. Experiments reveal that the proposed ABLNet has better feedback performance with adapting different lengths of input CSI. Meanwhile, ABLNet with FBCU can improve the model robustness and maintains the SGCS performance. The designed BNA algorithm can effectively stabilize the SGCS performance for every input CSI with fewer number of feedback bits. Finally, separate training delivers comparable feedback performance to joint training model, and can reduce the complexity of feedback model.

\bibliographystyle{IEEEtran}
\bibliography{ref2}


\end{document}